\documentclass[a4paper,fleqn,usenatbib]{mnras}

\usepackage{amsmath}
\usepackage{amsfonts}
\usepackage{amsbsy}
\usepackage{amssymb}
\usepackage{amsbsy}

\usepackage[T1]{fontenc}
\usepackage{ae,aecompl}

\usepackage{graphicx}
\usepackage{gensymb}
\usepackage{float}
\usepackage[utf8]{inputenc}
\usepackage{tensor}
\usepackage{bm}
\usepackage{enumerate}

\DeclareMathAlphabet\mathbfcal{OMS}{cmsy}{b}{n}

\defcitealias{riols17c}{RL18a}
\defcitealias{riols18}{RL18b}

\title[]{Gravitoturbulent dynamos in astrophysical discs}
\author[]{
A. Riols,$^{1}$ H. Latter $^{2}$
\\
$^{1}$Univ. Grenoble Alpes, CNRS, Institut de Planétologie et d’Astrophysique de Grenoble (IPAG), F-38000, Grenoble, France\\
$^{2}$DAMTP, University of Cambridge, Centre for Mathematical Sciences,
Wilberforce Road, Cambridge CB3 0WA, UK.}

\date{Accepted XXX. Received YYY; in original form ZZZ}

\pubyear{2017}

\begin{document}
\label{firstpage}
\pagerange{\pageref{firstpage}--\pageref{lastpage}}
\maketitle

\begin{abstract}
The origin of large-scale and coherent magnetic fields in
astrophysical discs is an important and long standing problem.
Researchers commonly appeal to a turbulent dynamo, sustained by the magneto-rotational
instability (MRI), to supply the large-scale field. 
But research over the last decade in particular
has demonstrated that various non-ideal MHD effects can impede or
extinguish the MRI, especially in
protoplanetary disks. In this paper we propose a new scenario, by
which the magnetic field is generated and sustained via the
gravitational instability (GI). We use 3D stratified shearing box
simulations to characterise the dynamo and find that it works at low
magnetic Reynolds number (from unity to $\sim 100$) for a wide range
of cooling times and boundary conditions. The process is kinematic,
with a relatively fast growth rate ($\lesssim 0.1\Omega$), and it
shares some properties of mean field dynamos. The magnetic field is generated via
the combination of differential rotation and spiral density waves,
the latter providing compressible horizontal motions and large-scale vertical
rolls.  
At greater magnetic Reynolds numbers the build up of large-scale
field is diminished and instead small-scale
structures emerge from the breakdown of twisted flux ropes.
We propose that GI may be key to the dynamo engine not 
only in young protoplanetary discs but also in some AGN and galaxies. 
\end{abstract}




\section{Introduction}
Magnetic fields appear in almost all astrophysical discs, from those surrounding
newly born stars, active galactic nuclei (AGN),
dwarf novae and low-mass X-ray binaries, to spiral galaxies. They are thought  to
guide the evolution of these objects via
several processes, such as accretion, outbursts, winds, and turbulence
\citep{balbus03,wardle07,armitage15}.  In several protoplanetary (PP) discs,
indirect
observations, based on polarized synchrotron or dust emission, have
detected magnetic fields with various strengths and morphologies
 \citep{carrasco10,stephens14,goddi17}. 
In fact, in the bulk of the FU-Ori disc, fields have been observed
directly, 
using the high-resolution
spectropolarimeter ESPaDOnS \citep{donati05}.
On the other hand, magnetic fields appear in many galactic discs, with a horizontal 
bisymmetric-spiral component \citep[e.g.  M51,][]{fletcher11} and in  
our own galaxy, with a vertical structure recently determined by the
Planck satellite using the cosmic background polarization.
Fields in AGN and mass-transferring binaries have not been detected directly,
but their existence can be inferred from the magnetised jets launched
from their associated disks. 

A fundamental goal is to determine the origin of these magnetic fields
and their sustenance over the disc's life. In the case of PP discs
(the main focus of our paper), it is possible
that the fields issue from the central protostar or are inherited
from the primordial nebulae. If the latter, then the
transport and amplification of magnetic field through the disk are
key, and these processes remain the focus of current research
\citep{guilet13,zhu18}. Another possibility is that
the field is produced in-situ by a turbulent dynamo. This idea,
originally suggested by \citet{pudritz81}, was bolstered
by the observations of \citet{donati05} which revealed a complex
magnetic topology, with both filamentary and toroidal structures,
possibly generated by motions internal to the disc. 
Such an in-situ process is appealing because of its generality: PP
disks need not rely on the particularities of
any external magnetic source; it might also generalise to AGN and
galactic disks.

For more than two decades, the magnetorotational instability (MRI) has
been proposed as a potential source of turbulence and magnetic fields 
in these objects \citep{balbus91,hawley95, branden95, balbus08}, via a 
subcritical, nonlinear, and cyclic dynamo. That said, in PP
discs, this process has been shown to work reliably only in their very inner radii ($\lesssim 0.1-1$ AU), where ionization sources are
strong enough to permit the MRI. Further out,
non-ideal MHD effects (Ohmic and ambipolar diffusion plus the Hall
effect) tend to weaken or quench the MRI \citep{wardle99,sano2000,fleming00,
kunz04,wardle12,bai13,lesur14, bai15}.
So too in AGN; close to the supermassive black
hole the disc is well coupled to the magnetic field, but outward
of 0.01 pc the state of the gas is far less clear, as it depends
on uncertain non-thermal ionisation sources and the degree
of disk warping \citep[e.g.][]{menou01}.
Putting ionisation aside, several numerical studies have suggested that the MRI dynamo is suppressed in many of these settings because of their low magnetic
Prandtl number \citep[ratio between  
the microscopic viscosity and resistivity; see][for more details]{fromang07b,balbus08,kapyla11,meheut15,riols16c}. 

An alternative to the MRI is growth of magnetic fields via other
disc instabilities. Perhaps the most efficient and powerful,
after the MRI,  is the gravitational instability \citep[GI; for the PP
disk context see reviews by][]{durisen07,kratter16}. Discs
are usually susceptible to
 the GI in their outer regions, provided that
\begin{equation}
Q=\dfrac{c_s\Omega}{\pi G \Sigma_0}\lesssim 1
\label{mass_eq}
\end{equation} 
where $Q$ is  the Toomre parameter \citep{toomre64}, $c_s$ is the
sound speed, $\Omega$
 the orbital frequency, and $\Sigma_0$ the background surface
 density. 
It is believed  that 50\% of class 0  and 10-20\% of class I PP discs might be unstable to GI 
\citep{tobin13, mann15}.  In particular, the detection of spiral arms
and streamers, recently imaged at NIR wavelengths in Ophiuchi and FU Ori systems,
supports the
idea  that their discs are gravitationally unstable
\citep{christiaens14,liu16,dong16}. 
Because they are
thin, AGN are also susceptible to GI in the regions beyond 0.01 pc and
might build up magnetic fields through this process \citep{menou01,goodman03, lodato07}.
The instability, of course, features also in galactic discs (perhaps its
most famous venue), shaping the large-scale 
spiral and clumpy structures where stars form \citep{goldreich65,wang94,kim2001}.

The ability of GI to amplify seed magnetic fields has been pointed out
in the context of star formation \citep[i.e during the collapse of
self-gravitating dense cloud, see][]{pudritz89,federrath11}. However,
the viability of a GI-reliant dynamo process is relatively unexplored in
the accretion disc context. 
The recent shearing
box simulations of \citet[][hereafter \citetalias{riols17c}]{riols17c}
showed that such a dynamo does exist
 and can sustain magnetic fields to near
equipartition values, especially in the regime of efficient cooling; 
 in addition, GI impedes the (zero-net flux) MRI in this
regime. However, most of the simulations were run in the ``ideal limit"
(i.e.\ without explicit diffusion), omitting all the relevant non-ideal
effects, such as resistivity, the Hall effect, and ambipolar diffusion,
each potentially important in the regions of PP discs where GI is active
\citep[see] [for instance]{simon15}.  Moreover, no theory was proposed
that could account for the properties of, and physical mechanism animating, 
the GI
dynamo.  This paper is a first in a series that aims to address these
issues and lay the foundations of a general theory.
Here, we only focus on ohmic resistivity,
so as to  more easily characterize the dynamo's fundamental nature: is it kinematic or nonlinear?
Fast or slow? Small or large scale?.  We stress that the present
study is a preliminary and 
 necessary step before including other non-ideal physics. {However, we
point out that both ambipolar diffusion and the Hall effect
are necessarily absent in the kinematic phase of any dynamo,
and so during this initial phase our results are broadly applicable.}

To that end, we perform 3D MHD stratified shearing box simulations
with self-gravity using a modified version of  the PLUTO code. 
Simulations maintain a quasi-steady energy balance via inclusion
of a simple linear cooling law, characterised by a uniform and
constant timescale $\tau_c$, and enforce a zero-net vertical
magnetic flux. We explore a range of magnetic Reynolds numbers, from 1 to
about 500, defined with respect to the sound speed and disc
scaleheight.
{It is important to note that the MRI
is excluded in this regime \citep{simon12}, and so any magnetic field
generation must issue from GI-turbulent motions}.  Given the size of the box necessary
to capture GI,  it is not yet possible to probe the regime of Rm
larger than a thousand, typically found in the inner ($\lesssim$0.1
AU) and external  regions ($\gtrsim$ 20 AU) of older PP disc, and in other
environments like AGN. 

We find that the GI-dynamo is inherently kinematic and
linear, with growth rates depending strongly on Rm. The dynamo process persists even in a very resistive plasma with typical $\text{Rm}\simeq 1$, and appears to be robust even when GI is weak. At lowish Rm ($\lesssim 100$), a very efficient dynamo, supported by the combined action of shear (the ``omega effect'') and
spiral wave motions, amplifies and organises the magnetic field into
large-scale spiral patterns. Essential ingredients in the dynamo are
the  vertical rolls associated with
spirals waves \citep[][hereafter
\citetalias{riols18}]{riols18} which allow the regeneration of the mean poloidal field from toroidal field, through a very unusual ``alpha" effect. At larger Rm ($\gtrsim 100$), the dynamo persists but saturates at a lower amplitude. In this regime the large-scale field ropes
 are twisted and break down into small-scale filamentary
structures with a concomitant alteration of the mean-field dynamo effect. 

The structure of the paper is as follows: first, in
Section \ref{sec_model}, we present the basic equations of the
problem and the numerical methods. In Section \ref{sec_threshold}  we study the dynamo growth rate and saturated state as a function of Rm. 
In Section \ref{sec_dynamo_scale}, we characterise the essence of the
dynamo process and its typical scales. In Section \ref{sec_spirals} we
investigate in more detail the role of spiral
waves in the generation of large-scale and also small-scale fields.
 We discuss in Section \ref{sec_discussion} the dependence of the dynamo on numerical details (boundary conditions, numerical resolution) and its applications to protoplanetary discs and galaxies. 
Finally, we state our conclusions in Section \ref{sec_conclusions}.

\section{Numerical model}
\label{sec_model}
\subsection{{Governing equations}}
We use the local Cartesian model of an accretion disc \citep[the
shearing sheet;][]{goldreich65,latter17} where the differential rotation is approximated locally
by a linear shear flow $-S \mathbf{e}_y$ and a uniform rotation rate $\boldsymbol{\Omega}=\Omega \, \mathbf{e}_z$, with $S=(3/2)\,\Omega$ for a Keplerian equilibrium. We denote $(x,y,z)$ as the radial, azimuthal and vertical directions respectively, and refer to the $(x,z)$ projections of vector fields as their poloidal components and to the $y$ component as their toroidal one. We assume that the gas orbiting around the central object is ideal, its pressure $P$ and density $\rho$ related by $\gamma P=\rho c_s^2$, where $c_s$ is the
sound speed and $\gamma=5/3$ the ratio of specific heats. In this paper,
we neglect molecular viscosity but will consider non-zero uniform
magnetic diffusivity $\eta$.{A uniform
  $\eta$ greatly simplifies the problem, although it is not
especially realistic. In PP discs, outside of 1 AU, the ionisation fraction (and consequently $\eta$) 
depends strongly on $z$, due to irradiation by cosmic, FUV, and X
rays. We assume for the moment that this variation
does not significantly influence the dynamo process. Future
simulations will explore a height dependent $\eta$, using realistic
profiles such as in \citet{simon15}}. 

The evolution of density ${\rho}$, total velocity {field} $\mathbf{v}$, magnetic field $\mathbf{B}$ and {total energy density $e_t=\dfrac{1}{2} \rho \mathbf{v}^2 + \mathbf{B}^2/2+ P/(\gamma-1)$ follows}:
\begin{equation}
\dfrac{\partial \rho}{\partial t}+\nabla\cdot \left(\rho \mathbf{v}\right)=0
\label{mass_eq}
\end{equation}
\begin{equation}
\frac{\partial{\mathbf{v}}}{\partial{t}}+\mathbf{v}\cdot\mathbf{\nabla
  v} +2\boldsymbol{\Omega}\times\mathbf{v} =-\nabla\Phi
  -\dfrac{\mathbf{\nabla}{P}}{\rho}+\dfrac{(\nabla\times \mathbf{B})\times\mathbf{B}}{\rho},
\label{ns_eq}
\end{equation}
\begin{equation}
\frac{\partial{\mathbf{B}}}{\partial{t}} =-\nabla\times \mathbf{E},
\label{magnetic_eq} 
\end{equation}
\begin{equation}
{\dfrac{\partial e_t}{\partial t}+\nabla\cdot \left [\left(e_t+P+\mathbf{B}^2/2)\mathbf{v}+\mathbf{E}\times \mathbf{B}\right )\right]
  =\rho \mathbf{v}\cdot \mathbf{\nabla \Phi}-\dfrac{P}{\tau_c},}
\label{heat_eq}
\end{equation}
{where $\mathbf{E}=-\mathbf{v}\times\mathbf{B}+\eta\,\mathbf{\nabla}\times \mathbf{B}$ is the electric field}. The total velocity field can be decomposed into a mean shear and a perturbation $\mathbf{u}$: 
\begin{equation}
\mathbf{v}=-S x\,  \mathbf{e}_y+\mathbf{u}.
\end{equation}
$\Phi$ is  the sum of the tidal gravitational potential induced by the
central object in the local frame $\Phi_c=\frac{1}{2}\Omega^2
z^2-\frac{3}{2}\Omega^2\,x^2$  and the gravitational potential
$\Phi_s$ induced by the disc itself. The latter obeys the Poisson equation: 
\begin{equation}
\mathbf{\nabla}^2\Phi_s = 4\pi G\rho.
\label{poisson_eq}
\end{equation}
Cooling in the {total energy equation} (\ref{heat_eq})
is a linear function of $P$ with a typical timescale $\tau_c$ referred
to as the `cooling time'. This prescription is not especially
realistic but allows us to simplify the problem as much as possible.
We also neglect thermal conductivity. 

Finally, $\Omega^{-1}=1$ defines our unit of time and $H_0=1$ our unit of length. $H_0$ is the standard hydrostatic disc scale height defined as the ratio $c_{s_0}/\Omega$ where $c_{s_0}$ is the sound speed in the midplane of a hydrostatic disc in the limit $Q\rightarrow \infty$. To characterize the importance of ohmic dissipation in the induction equation, we introduce the magnetic Reynolds number: 
\begin{equation}
\text{Rm}=\dfrac{c_{s_0}  H_0}{\eta}.
\end{equation}

\subsection{Numerical methods}

Our computational approach is identical to that used in
\citetalias{riols17c}. Simulations are performed with the Godunov-based PLUTO code, adapted to highly compressible flow \citep{mignone2007}, in the shearing box framework. The box has a finite domain of  size $(L_x,L_y,L_z)$, discretized on a mesh of $(N_X,N_Y,N_Z)$ grid points. The numerical scheme uses a conservative finite-volume method that solves the approximate Riemann problem at each inter-cell
boundary. It conserves quantities like mass, momentum,
and total energy across discontinuities. The Riemann problem is handled by the HLLD solver, suitable for MHD. An orbital advection algorithm is used to increase the computational speed and reduce numerical dissipation.  Finally, the divergence of $\mathbf{B}$
is forced to 0 by the 
constrained-transport algorithm of PLUTO. 

To calculate the 3D self-gravitating potential, we employ the numerical methods detailed in \citet{riols17b} and \citetalias{riols17c}. We tested the stratified disc equilibria, as well as the linear stability of these equilibria, to ensure that our 
implementation is correct (see appendices in \citet{riols17b}). 

 The boundary conditions are periodic in $y$ and shear-periodic in $x$. In the vertical direction,  we use a standard outflow condition for the velocity field and assume a hydrostatic
balance in the ghost cells for pressure, taking into account the 
large-scale vertical component of  self-gravity (averaged in $x$ and $y$).
For the magnetic field, we usually impose  $B_x=B_y=0$ and $dB_z/dz=0$, as
in several previous MRI simulations
\citep{branden95,gressel10,oishi11,kapyla11}. They allow the mean
horizontal magnetic field (or total flux)  to vary, even
when the field has initially zero net $B_x$ or $B_y$, and 
thus a potential concern
is that a mean-field dynamo might be
artificially sustained. That being said, previous studies
indicate  that the large-scale magnetic field behaves
similarly when open or closed boundary conditions are employed.  In
any case, in Section \ref{boudary}, we simulate the dynamo in a limited set of runs with
periodic boundaries, where this is no issue.
(See also the discussion
in Section \ref{emf_terms}.) Finally, the boundary conditions for the
self-gravity potential, 
in Fourier space, are:
 \begin{equation}
\dfrac{d}{dz} \Phi_{k_x,k_y} (\pm {L_z}/{2}) = \mp k\Phi_{k_x,k_y} (\pm {L_z}/{2}).
\end{equation} 
where $\Phi_{k_x,k_y}$ is the horizontal Fourier component of the potential and $k_x$, $k_y$ are the radial, azimuthal wavenumbers and $k=\sqrt{k_x^2+k_y^2}$. This condition is an approximation of the Poisson equation in the limit of low density. 

Lastly, we enforce  a density floor of $10^{-4}\, \Sigma/H_0$ which
prevents the timesteps getting too small due to evacuated regions near
the vertical boundaries. Mass is replenished near the midplane so that
the total mass in  the box is maintained constant.  
We checked that the mass injected at each orbital period is negligible
compared to the total mass (less than 1\% per orbit). 

\subsection{Simulation setup and parameters}
\label{sim_setup}

The large scale waves excited by GI have 
radial lengthscales $\lambda \gtrsim H\,Q$. In order to capture these waves, while affording acceptable
resolution, we use a box of intermediate horizontal size $L_x=L_y=20 \, H_0$. The vertical domain of the box
spans $-3\,H_0$ to $3\,H_0$.  For runs with Rm $\leq 200$, we use a
resolution of $N_X=256,N_Y=256$ and $N_Z=96$.
 This resolution is enough to capture the smallest scales at which
 magnetic energy is (physically) dissipated.
 For simulations without explicit resistivity or $\text{Rm}>200$, we
double the resolution in $x$ and $y$ ($N_X=512,N_Y=512$ and $N_Z=128$).  

Throughout we use a fixed heat
capacity ratio $\gamma=5/3$ and consider a zero-net vertical magnetic
flux.  To run pure hydrodynamic GI simulations,  we start from a
polytropic vertical density equilibrium, computed with a Toomre Q of
order 1. The calculation of such an equilibrium is detailed in the
appendix  of \citet{riols17b}. Non-axisymmetric density and velocity
perturbations of finite amplitude are injected to trigger the
turbulent state. For the MHD runs, the initialization of $\mathbf{B}$
will be detailed in the corresponding sections. 

\subsection{Diagnostics}
\begin{figure*}
\centering
\includegraphics[width=\textwidth]{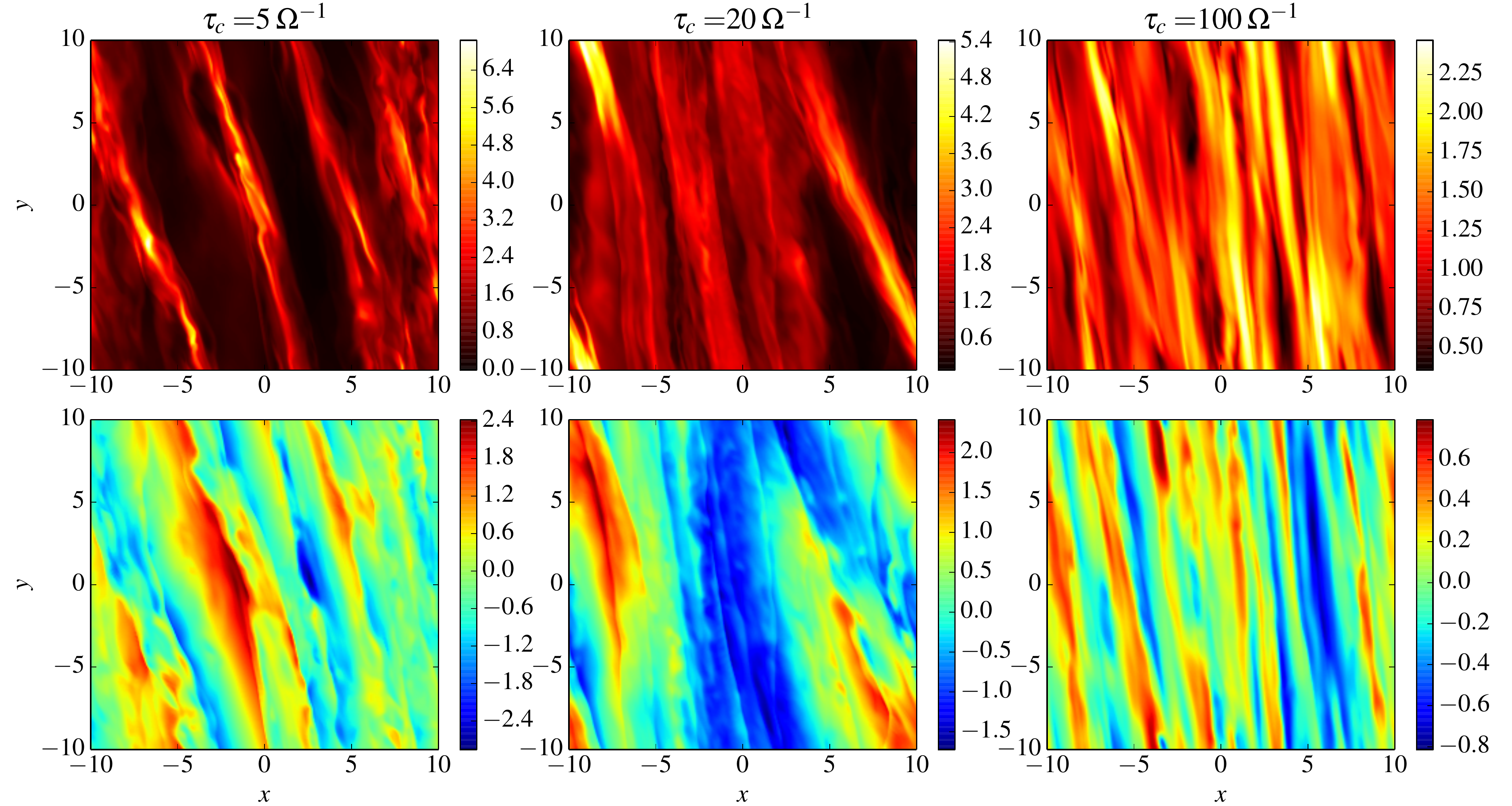}
 \caption{Hydrodynamic gravito-turbulent states used as initial conditions for the dynamo problem. The colourmaps represent gas density $\rho$ (top) and radial velocity $u_x$ (bottom)  in the midplane $z=0$.  From left to right, $\tau_c=5, 20$ and $100\,\Omega^{-1}$.  }
\label{fig_GIhydro}
 \end{figure*} 
\subsubsection{Averages}
To analyse the statistical behaviour of the turbulent flow, we define the standard and weighted box averages
\begin{equation}
\left<X \right>=\frac{1}{V}\int_V X\,\, dV,  \quad \text{and} \quad \left<X \right>_W=\frac{1}{V \left<\rho \right>}\int_V \rho  X\,\, dV, 
\end{equation}
respectively,
where $V=L_x L_y L_z$ is the volume of the box. We also define the horizontally averaged vertical profile of a dependent variable: 
\begin{equation}
\overline{X}(z)=\dfrac{1}{L_xL_y} \int\int X\,\, dxdy.
\end{equation}
An important quantity is the average 2D Toomre parameter defined by
\begin{equation}
Q_W=\dfrac{\left\langle c_s\right\rangle_W\Omega}{\pi G  \Sigma },
\end{equation}
where  $\Sigma=L_z \left<\rho \right>$ is the average
surface density of the disc. Another quantity that characterises the turbulent dynamics is the coefficient $\alpha$ which measures the angular momentum transport efficiency. This quantity is the sum of the total stress (which includes gravitational $G_{xy}$, Reynolds $H_{xy}$ and Maxwell stresses $M_{xy}$) divided by the box average pressure:
\begin{align}
\label{def_alpha}
\alpha=\dfrac{\left\langle
  H_{xy}+G_{xy}+M_{xy} \right\rangle}{\langle P\rangle},
\end{align} 
where
\begin{align*}
H_{xy}=\rho u_xu_y,
\quad G_{xy}=\dfrac{1}{4\pi G}\dfrac{\partial \Phi}{\partial
  x}\dfrac{\partial \Phi}{\partial y} \quad \text{and} \quad M_{xy}=-B_xB_y.
\end{align*}
Finally, we denote by $E_c=\frac{1}{2}\langle\rho \mathbf{u}^2\rangle$  and $E_m=\frac{1}{2}\langle\mathbf{B}^2\rangle$ the box averaged kinetic and magnetic energies.

\subsubsection{Fourier modes}

To analyse the structure of the magnetic field,  we use the 2D Fourier
decomposition of these fields in $x$ and $y$. The Fourier modes are
calculated as in \citet{riols17b} (Section 2.5.2). We denote by $k_x$
and $k_y$ the radial and azimuthal Eulerian wavenumbers. In this
mathematical representation,  fields are a sum of a \emph{mean} ($k_x=0$ and
$k_y=0$) mode, axisymmetric modes with $k_y=0$ ($k_x\neq0$), and
non-axisymmetric modes with $k_y\neq0$ (commonly referred as ``shearing
waves").
 A large-scale wake or spiral wave can be viewed as the result of
the constructive interference of multiple
shearing waves of different $k_x$ but the same,
fundamental, azimuthal wavenumber $k_y=2 \pi/L_y$  \citep[see][]{ogilvie02}

\subsubsection{Mean magnetic field and dynamo equation}

In Section \ref{sec_dynamo_scale}, we investigate the ability of GI to
sustain a large-scale dynamo throughout the domain. A key quantity is
the mean magnetic field  $\overline{\mathbf{B}}(z)$, averaged in the
$x$ and $y$ directions, corresponding to the $k_x=0$ and $k_y=0$   
Fourier component of $\mathbf{B}$. The equations governing the horizontally-averaged radial and toroidal fields are: 
\begin{equation}
\label{eq_Bxmean}
\dfrac{\partial \overline{B}_{x}}{\partial t} =- \dfrac{\partial \mathcal{\overline{E}}_{y}}{\partial z} + \dfrac{1}{\text{Rm}}  \dfrac{\partial^2 {\overline{B}}_{x}}{\partial z^2} 
\end{equation}
\begin{equation}
\label{eq_Bymean}
\dfrac{\partial \overline{B}_{y}}{\partial t} = \ - S \overline{{B}}_{x}  + \dfrac{\partial \mathcal{\overline{E}}_{x}}{\partial z} + \dfrac{1}{\text{Rm}}  \dfrac{\partial^2 {\overline{B}}_{y}}{\partial z^2} 
\end{equation}
with $ \overline{\mathbfcal{E}}(z)= \overline{\mathbf{u} \times
  \mathbf{B}}$ the horizontally-averaged electromotive force (EMF). If
we denote $\mathbf{\tilde{b}}=\mathbf{B}-\overline{\mathbf{B}}$ and
$\tilde{\mathbf{u}}=\mathbf{u}-\overline{\mathbf{u}}$,  i.e.\
perturbations in the magnetic and velocity fields, then we have 
\begin{equation}
\overline{\mathbfcal{E}}(z)= \overline{\mathbf{u}} \times \overline{\mathbf{B}}+ \overline{\mathbf{\tilde{u}} \times \mathbf{\tilde{b}}}
\end{equation}
The first ``laminar" term is due to the mean
flow, while the second ``turbulent' term issues from the
correlation between velocity and magnetic perturbations. We recall
that the mean radial and toroidal fields in the box, respectively
$\langle B_x \rangle$ and $\langle B_y \rangle$ (equal to the
$z$-average of $\overline{{B}}_{x}(z)$ and $\overline{{B}}_{x}(z)$),
are free to vary.

\section{GI-dynamo threshold and saturation}
\label{sec_threshold}

In the regime of moderate cooling ($\tau_c \gtrsim 3\, \Omega^{-1})$
the GI breaks down into a disordered state, often referred to as
``gravito-turbulence", the properties of which have prompted
considerable study
\citep{Gammie2001,rice03,Pdkooper2012,shi2014,hirose17,riols17b}.
 The aim of this section
is to determine, for a given Rm and by using numerical simulations,
the ability of gravito-turbulent flows to  generate magnetic
fields. We calculate, in particular, the growth rates and saturated states
of the associated dynamo as a function of Rm. We stress that the large
resistivity in our simulations 
prohibits excitation of the MRI, so any field growth cannot be
attributed to it.

\begin{center}
\begin{table*}    
\centering 
\begin{tabular}{c c c c c c c c c c}          
\hline                        
Run & Resolution & Time (in $\Omega^{-1}$) & $\tau_c$ & $Q_W$ & $E_c$/$\langle P \rangle$ & $\langle H_{xy} \rangle$/$\langle P \rangle$ & $\langle G_{xy}\rangle$/$\langle P \rangle$ & $\alpha$ &  $\alpha_{th}=1/(\Omega\tau_c)$\\    
\hline  
	SGhydro-5 	& $256\times 256\times 96$ & 100 & 5 & 1.239 & 0.67 & 0.063 & 0.117 & 0.18 & 0.2\\
    SGhydro-20 &  $256\times 256\times 96$ & 200  & 20 & 1.194 &  0.191 & 0.0132 & 0.0345 & 0.048 & 0.05\\      
    SGhydro-100 & $256\times 256\times 96$ &  400 & 100  & 1.29 & 0.088 & 0.0032 & 0.00804 & 0.011 &  0.01\\
    \hline
\\                              
\end{tabular}  
\vspace{0.5cm}
\caption{Parameters and properties of hydrodynamical runs in a box of
  $L_x=L_y=20 H_0$. The third columm indicates the time of the
  simulation over which quantities have been averaged. $Q_W$ is the
  (density weighted) average Toomre parameter, $E_c/\langle P \rangle$
  is the ratio of box-averaged kinetic energy over pressure, $H_{xy}$,
  $G_{xy}$ and $\alpha$ are the averaged Reynolds, gravitational
  stress, and angular momentum transport, respectively.}  
\vspace{0.5cm}
\label{table1}
\end{table*}  
\end{center}

\subsection{Hydrodynamic flows}

Numerically, the usual way to study dynamos in turbulent flows is to
start from the hydrodynamic state of
fully-developed turbulence, and then introduce a magnetic seed to see
whether it grows or decays over time. To prepare
the initial states, we ran a set of pure hydrodynamic GI
simulations, as in \cite{riols17b}, with
moderate resolution (13 points per $H_0$ in the horizontal directions
and 16 points in the vertical). This resolution allows us to resolve
magnetic Reynolds number up to 200 (see Section \ref{sim_setup}). 

As described in \cite{riols17b} the
strength and saturation of the gravito-turbulence is set by the
cooling time $\tau_c$. 
In particular, the time-averaged stress to
pressure ratio follows the \citet{Gammie2001} relation:  
\begin{equation}
\label{gammie_eq}
\alpha\simeq\dfrac{1}{q\Omega (\gamma-1)\tau_c} = \dfrac{1}{\Omega \tau_c}  \quad (\gamma=5/3, \, q=3/2) 
\end{equation}
Three different $\tau_c$ have been considered: $\tau_c=5\,
\Omega^{-1}$ (strong turbulence, close to the fragmentation
threshold),
 $\tau_c=20\, \Omega^{-1}$ (moderate turbulence) and $\tau_c=100\,
 \Omega^{-1}$ (weak turbulence). 

Table \ref{table1} shows some box and time-averaged quantities for the
three different hydrodynamical simulations. 
While the Toomre parameter varies slightly between simulations,
kinetic energy, Reynolds and 
gravitational stress increase significantly when $\tau_c$ decreases. 
We checked also that the Gammie relation (Eq.~\ref{gammie_eq}) holds
in the three cases. 

Figure \ref{fig_GIhydro} shows the density and radial velocity of
the turbulent state associated with each $\tau_c$.  These representative snapshots
are chosen at a random time and serve as initial conditions for
the dynamo simulations in the next sections. For
$\tau_c=5\,\Omega^{-1}$ and $\tau_c=20\,\Omega^{-1}$, the turbulence
is supersonic, highly compressible, and characterized by large-scale
 spiral density waves, particularly strong in the case $\tau_c=5 \,
\Omega^{-1}$. Small-scale motions, potentially driven by a parametric
instability \citep[see][]{riols17b} faintly distort the spiral waves at
higher altitude, around $z\simeq H_0$. They are also marginally visible
in the midplane  (Fig~.\ref{fig_GIhydro}) but with weaker
amplitude. Note that the resolution used here allows us to capture only
the largest scales of the parametric instability.  In
the case $\tau_c=100 \,\Omega^{-1}$, the flow looks very different:
the turbulence is subsonic, the waves are fainter, smoother and
thinner, and their pitch angle much smaller.  

 \begin{figure}
\centering
\includegraphics[width=\columnwidth]{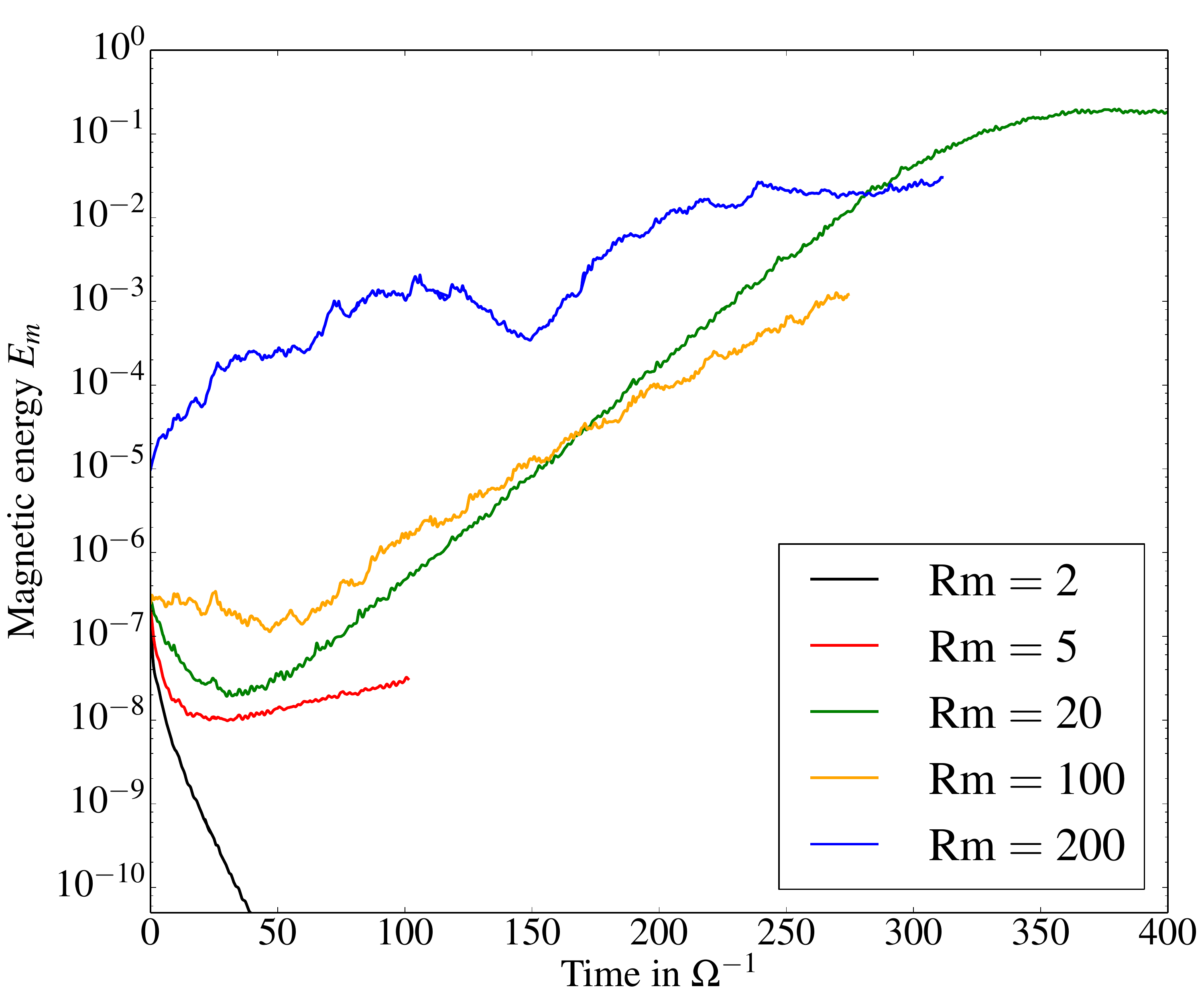}
 \caption{Time evolution of the  averaged magnetic energy for
   different Rm and for $\tau_c=20\,\Omega^{-1}$. The initial
   condition
 corresponds to the hydrodynamic state shown in Fig.~\ref{fig_GIhydro} (second column). }
\label{fig_Em}
 \end{figure}
 \begin{figure*}
\centering
\includegraphics[width=0.9\textwidth]{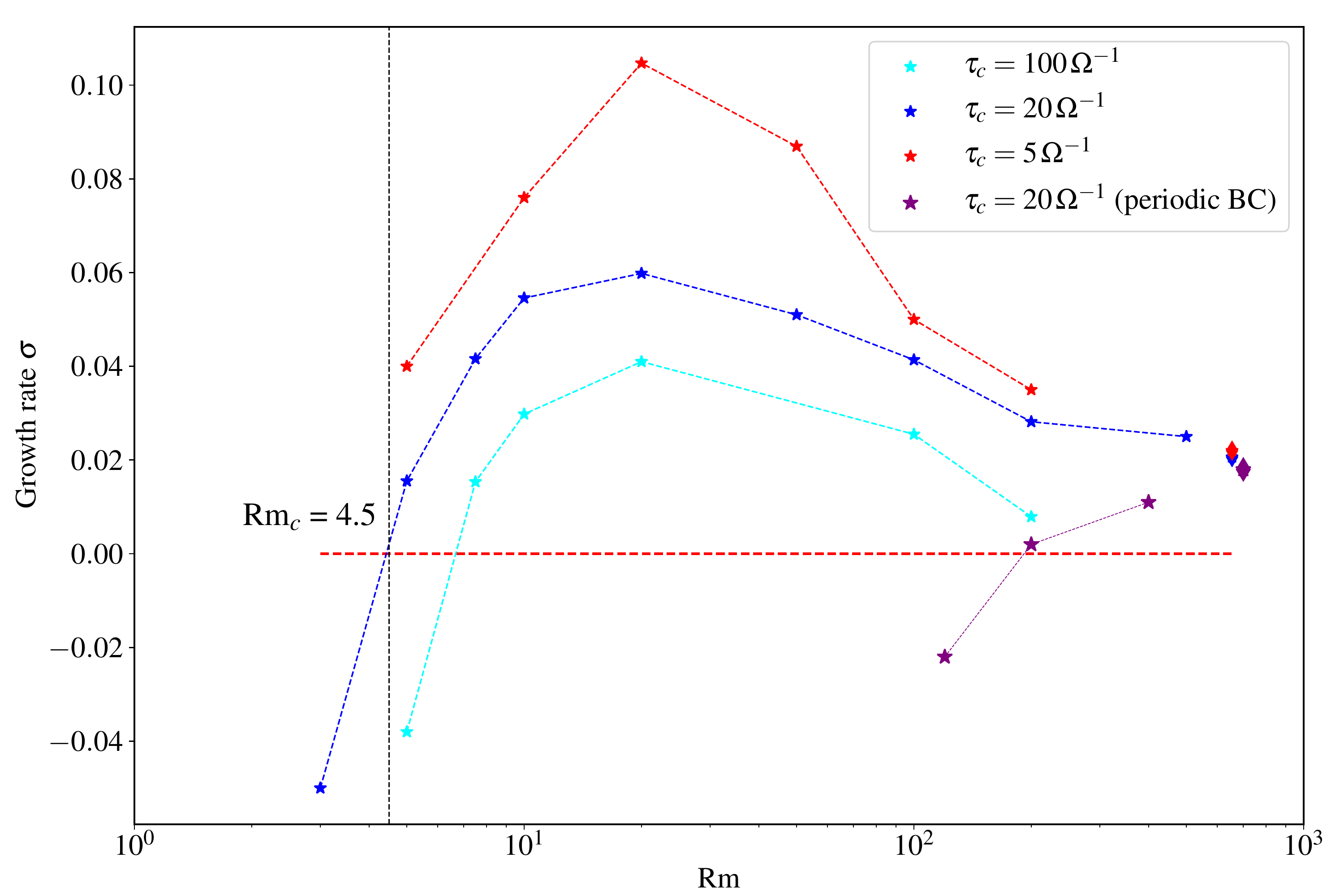}
 \caption{Linear growth rates of the magnetic energy as a function of
   Rm for 3 different cooling times, $\tau_c=5 \,\Omega^{-1}$ (red
   curve), $\tau_c=20 \,\Omega^{-1}$ (blue curve) and $\tau_c=100
   \,\Omega^{-1}$ (cyan curve).  The purple curve is for $\tau_c=20
   \,\Omega^{-1}$ but with periodic boundary conditions.  For a given
   $\tau_c$, the initial condition and initial seed magnetic field are
   identical for all Rm (the initial amplitude is different in the
   case $\tau_c=20\Omega^{-1}$, $\text{Rm}=200$ only).  The blue and
   red diamond markers  indicate the growth rate of magnetic energy
   obtained in the simulations SGMRI-20 and SGMRI-5  of
   \citetalias{riols17c}, respectively, without explicit resistivity and grid $\text{Rm} \sim 660.$}
\label{fig_growthrate}
 \end{figure*} 

\subsection{Kinematic regime and growth rates}

Having constructed the hydrodynamic GI turbulent states, we now
investigate their amplification of magnetic fields. For a
given $\tau_c$,  we run a series of MHD simulations at different Rm. 
The initial condition corresponds to the
state described in Fig.~\ref{fig_GIhydro}.  The magnetic seed introduced at the
beginning of the simulations is zero-net flux, toroidal, and has a sinusoidal
dependence in $z$.  In all cases (except for $\text{Rm}=200$), its
amplitude is fixed to $10^{-3}$, which corresponds to an initial $E_m=2.5\times
10^{-7}$.  This very small value ensures that
 the Lorentz force has little effect on the
gas dynamics during the first
few hundred $\Omega^{-1}$, at least on large scales, and thus any dynamo action is
purely kinematic.

We first analyse the intermediate cooling regime with
$\tau_c=20\,\Omega^{-1}$. Figure \ref{fig_Em} shows the time evolution
of the box-averaged magnetic energy $E_m$ for different Rm.   For
$\text{Rm}=2$, we found that  the magnetic energy falls to 0
with decay rate close to 0.14$\Omega$. For Rm between 5  and 100, $E_m$
evolves quasi-exponentially with a well-defined growth rate.  This is
particularly striking in the case of $\text{Rm}=20$ (green curve) for which a
clean exponential amplification is obtained for at least
$300\,\Omega^{-1}$. Saturation of magnetic energy occurs
after this point (a regime analysed in
Section \ref{saturation}).

 For the largest Rm shown in the figure
($\text{Rm}=200$, blue curve), we found that the magnetic field is
amplified with growth rate $\sigma \sim 0.028\Omega$, measured by fitting an
exponential curve to the numerical data.  Unlike the more resistive cases we explored, 
its evolution is more complex than a simple exponential
growth; the field exhibits intermittent bursts with long
periods of about $100 \,\Omega^{-1}$ and shorter periods of $\sim 10
\,\Omega^{-1}$.  We checked that the same behaviour is obtained when
$\text{Rm}=500$ with double the resolution. Unsurprisingly perhaps, it
resembles the dynamo identified by  \citetalias{riols17c} in the
limit of ideal MHD (without explicit resistivity). We are hence tempted to
conclude that the
simulations are only marginally resolved at this Rm.
 What we may be seeing is the grid (rather than explicit resistivity) beginning to control the dynamics.

We collect the growth rates obtained from all simulations with cooling time
$\tau_c=20\,\Omega^{-1}$ and plot their dependence on Rm in Figure
\ref{fig_growthrate} (blue curve).  To make a comparison with the
ideal limit, we also show in this diagram the growth rate measured in
the simulation SGMRI-20 of  \citetalias{riols17c} which had no
explicit resistivity  (blue diamond marker).  We deduce from this
figure that the critical threshold for the dynamo is $\text{Rm}
\gtrsim4- 5$.  Its growth rate is maximal for $\text{Rm}\simeq 20$
(with $\sigma_{max} \simeq 0.06\Omega$) and decreases at larger Rm.

In addition, we explore the dependence of the growth rates on
$\tau_c$, by repeating exactly the same procedure as described above,
for $\tau_c=100\,\Omega^{-1}$ and $\tau_c=5\,\Omega^{-1}$. The
corresponding growth rates are superimposed on
Fig.~\ref{fig_growthrate} to make direct comparison with
$\tau_c=20\,\Omega^{-1}$. In the case of inefficient cooling (cyan curve,
$\tau_c=100\,\Omega^{-1}$), the dynamo is still active and the
dependence on Rm is very similar to
$\tau_c=20\,\Omega^{-1}$. However, the growth rates are
lower and reduced by a factor 1.5$-$2. For efficient cooling (red
curve, $\tau_c=5\,\Omega^{-1}$), turbulent motions are stronger and
growth rates  are consequently larger. The maximum amplification occurs again for
$\text{Rm}\simeq 20$, with growth rate $\sigma \simeq 0.1\Omega$. In this
regime  magnetic fields are efficiently amplified so that they reach a
quasi-steady configuration within a few orbits.

To summarize, GI can amplify a magnetic field for a wide range of
$\tau_c$ and the dynamo persists at very small $\text{Rm} \gtrsim
4$. In addition to its kinematic and linear properties, we demonstrate that
the dynamo behaves like a ``slow" dynamo, with growth rates decreasing with increasing
Rm above a critical value.
The runs at double resolution with
$\text{Rm}=500$ and those without explicit resistivity (grid Rm
$\simeq H/\Delta_x^2\simeq $ 660)  suggest perhaps that the
growth rate decreases more slowly at higher Rm, indicative of a
different dynamo regime.
Despite
this trend, we emphasize that predicting the dynamo behaviour at
larger Rm or in the ideal limit remains extremely difficult. 
Testing the convergence of the dynamo growth rate with Rm
would require us to go beyond $\text{Rm} \sim 1000$, which is inaccessible
with our current resources. In any case, these results suggest that ohmic
diffusion is crucial for the maintenance of a powerful and efficient
dynamo. 

 \begin{figure}
\centering
\includegraphics[width=\columnwidth]{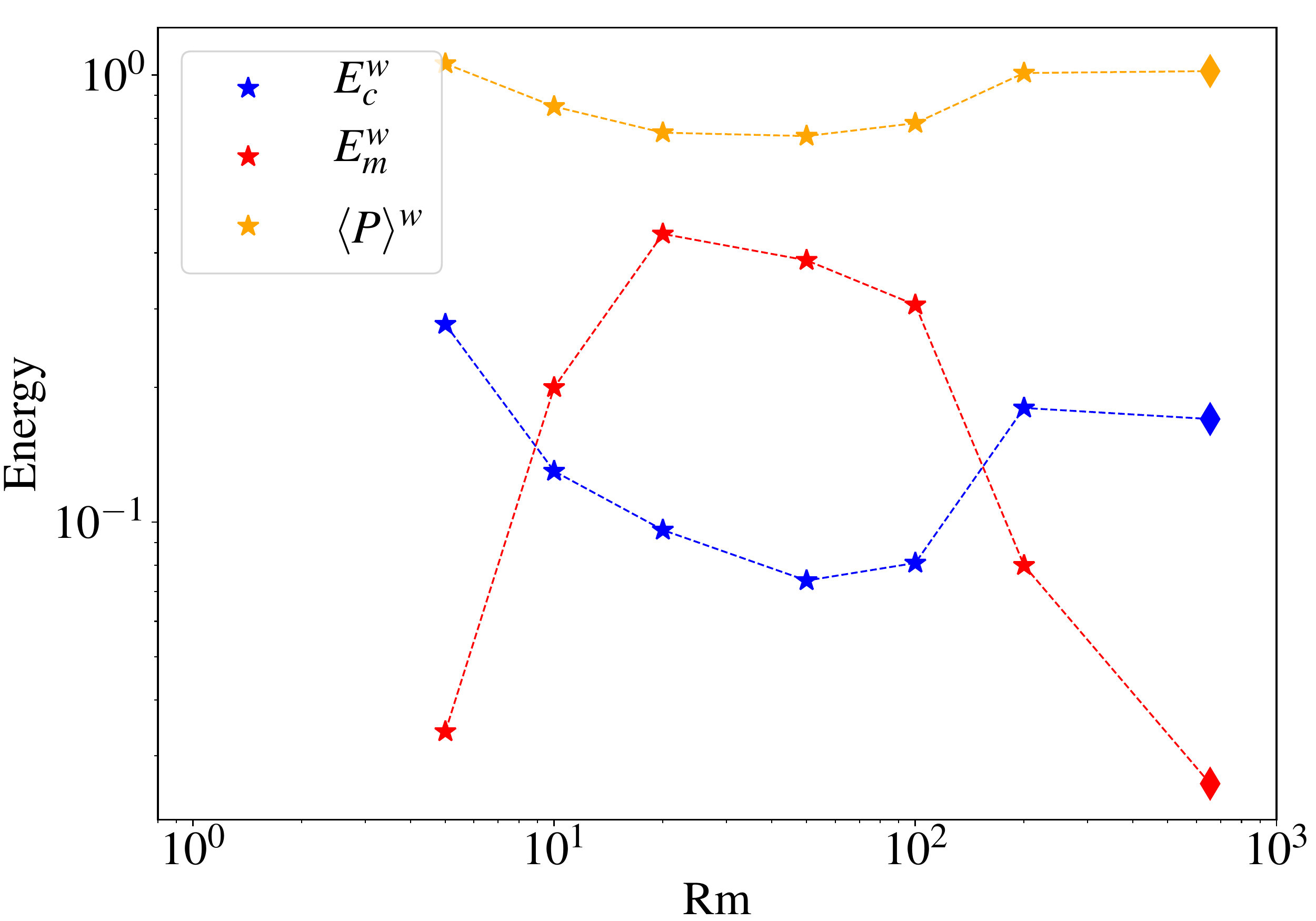}
 \caption{Pressure, kinetic and magnetic energy as a function of Rm
   for $\tau_c=20\, \Omega^{-1}$ during the quasi-steady nonlinear
   regime. 
 Quantities are weighted by the gas density and averaged over the box
 and. 
The diamond markers correspond to the simulation without explicit resistivity with grid $\text{Rm} \sim 660$. }
\label{fig_saturation}
 \end{figure} 
  \begin{figure}
\centering
\includegraphics[width=\columnwidth]{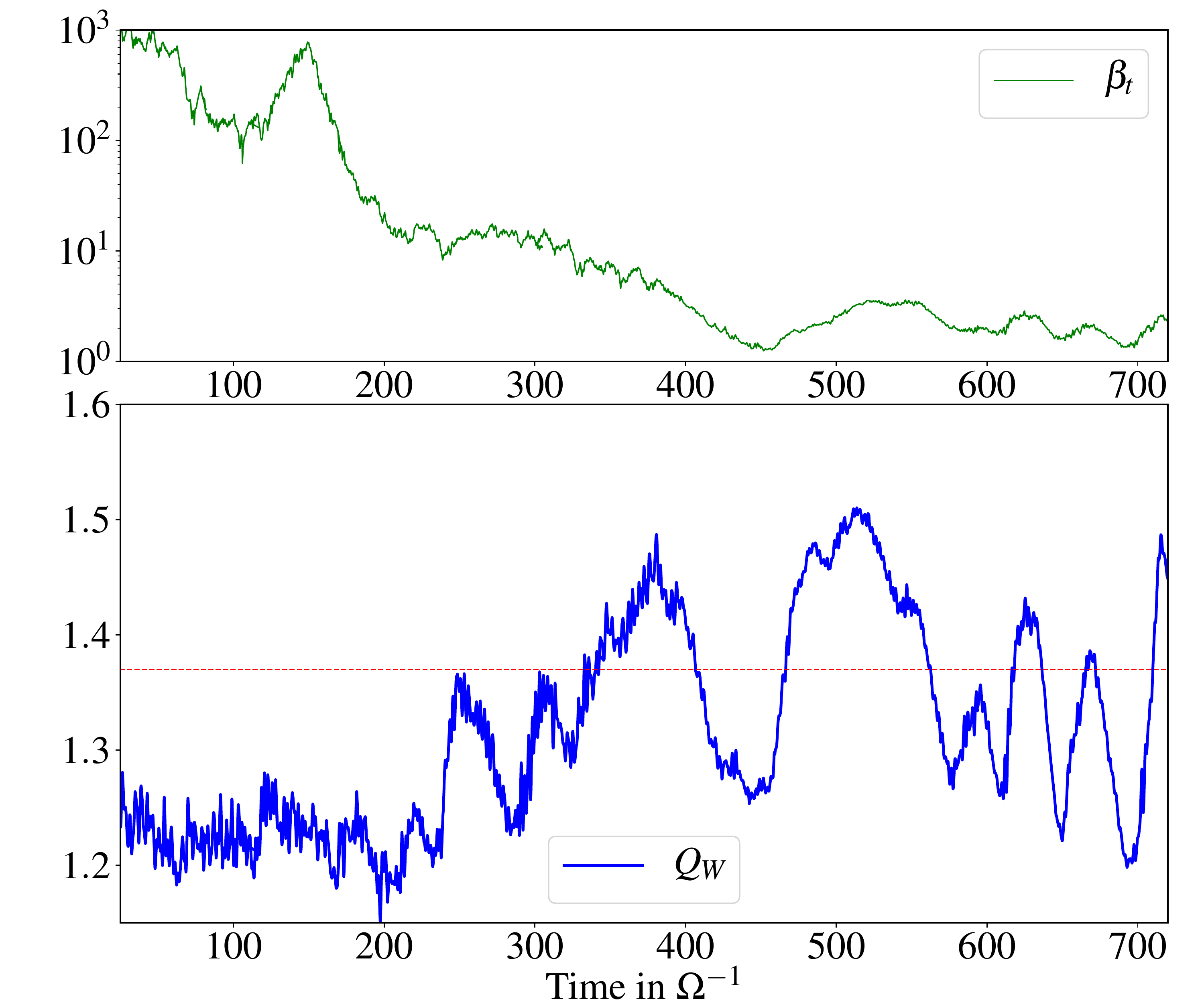}
 \caption{Time-evolution of turbulent beta plasma $\beta_t$ (top) and 
Toomre parameter (bottom) for $\text{Rm}=50$ and
$\tau_c=20\,\Omega^{-1}$. 
Quantities are averaged over the box with density weighting.}
\label{fig_toomre}
 \end{figure} 

\subsection{Saturation and nonlinear regime}
\label{saturation}

We now explore the non-linear regime of the dynamo,  i.e when the
Lorentz force provides feedback on the turbulent flow.  What is the
dependence of the saturated state on Rm and how strongly does
resistivity affect GI turbulence?

First, we plot in Fig.~\ref{fig_saturation} the box-averaged pressure,
kinetic and magnetic energy obtained in the saturated state (when the
magnetic field stops growing) as a function of Rm and for
$\tau_c=20\,\Omega^{-1}$.  Since the field concentrates within the
bulk of the disc (see Section \ref{fft_modes}), we weighted quantities
by the gas density during the averaging procedure.  We see that the
saturated magnetic energy  in the nonlinear regime depends on
the parameters in the same way as the kinematic growth rates (see Fig.~\ref{fig_growthrate} for
comparison).  $E_m$ is maximal for $\text{Rm}\ \sim 20$ and decreases
quite abruptly towards $\text{Rm}\gtrsim 200$. At its maximum, the dynamo
sustains strong magnetic fields with a plasma beta
$\beta_t=\langle P \rangle / E_m \sim 2$ and energy typically 5-6
times larger than the turbulent kinetic energy in the bulk of the
disc. This corresponds to a regime of super-equipartition fields, for
which we expect important feedback of the Lorentz force on the gas
flow. For Rm between 10 and 100, the kinetic energy is
reduced by a factor almost 3 compared to the ideal limit and by a
factor 6 compared to the unmagnetized gravitoturbulence.  We checked that the
gravitational stress is also reduced in this regime.  

In  \citetalias{riols17c}, we showed that ideal-MHD gravitoturbulence exhibits
a moderate drop in
kinetic energy in comparison to the purely hydrodynamical case, 
a property attributed to the non-linear suppression of 
large-scale structures, such as oscillatory epicyclic modes. In the low Rm
regime, the drop in kinetic energy is more dramatic
because of the stronger magnetic field. Additional effects, associated
with magnetic pressure and thermodynamics, may also contribute; 
for instance, since the kinetic to magnetic pressure ratio is of order unity
for Rm between 10 and 100, the linear stability and growth
rate of spiral GI modes will be altered. Indeed, the \emph{effective}
Toomre parameter, including magnetic pressure, is
\begin{equation}
Q_{\text{eff}}\approx Q_{\text{hydro}}  \sqrt{1+\dfrac{1}{\beta_t}}
\end{equation}
\citep{kim2001} which is $25\%$ greater than the hydrodynamic $Q$ in
the low Rm states. Moreover, magnetic fields provide an
additional reservoir of energy that can be directed into ohmic heating
and, consequently, an enhanced $Q$. 

To
study this effect, we plot in
Fig.~\ref{fig_toomre}  the evolution of the box-average Toomre
parameter $Q_W$ as a function of time for $\text{Rm}=50$. In the
initial state and during the kinematic phase, the magnetic field is
too small to influence the flow and $Q_W$ settles around its
hydrodynamical value ($Q_W\simeq 1.2$). In the final state, with
$\beta_t\approx 2$, the average Toomre parameter is $Q_W \simeq
1.37$. Therefore there is an increase of almost 15\% 
(and equivalently of the average sound speed in the midplane). Note
however that the elevation of disc temperature remains moderate, despite the strong
magnetization. In particular, it is relatively small compared to that
measured in a 2D disk \citep{riols16a}. Nonetheless, by
combining the effect of magnetic pressure and ohmic dissipation, the
effective $Q$ in the low Rm regime can be 40\% higher than in pure
hydrodynamics, leading to fainter spiral waves and a depression of
GI activity.

Interestingly, $Q$ exhibits notable oscillations when the
magnetization becomes important (at $t=300\, \Omega^{-1}$). These
oscillations must reflect the back-reaction of the field on the flow.
They probably originate from a thermal cycle similar to that described
in \citetalias{riols17c} (Section 5.4), in which  (a) GI amplifies
$\mathbf{B}$, (b) magnetic energy is dissipated into heat and enhances
$Q$, (c)  GI or spiral activity dies because of the high $Q$ and strong
magnetic tension, and finally (d) in the absence of vigorous GI, $\mathbf{B}$ and $Q$
return to their original value. This scenario is
consistent with the fact that the variations of  $Q$ and $\beta_t$ in
Fig.~\ref{fig_toomre} are correlated.

The dependence of saturated state on cooling times was
explored in \citetalias{riols17c}  in the ideal limit.  In the case of
finite resistivity, we find that magnetic energy also peaks around
$\text{Rm}=20$ for $\tau_c=100\,\Omega^{-1}$. However for
$\tau_c=5\Omega^{-1}$, simulations have not been run for sufficiently
long time to be statistically meaningful. The reason is that most
stalled after a few tens of an orbit because of transient fragments
that considerably reduce the time step.

\begin{figure*}
\centering
\includegraphics[width=0.497\textwidth]{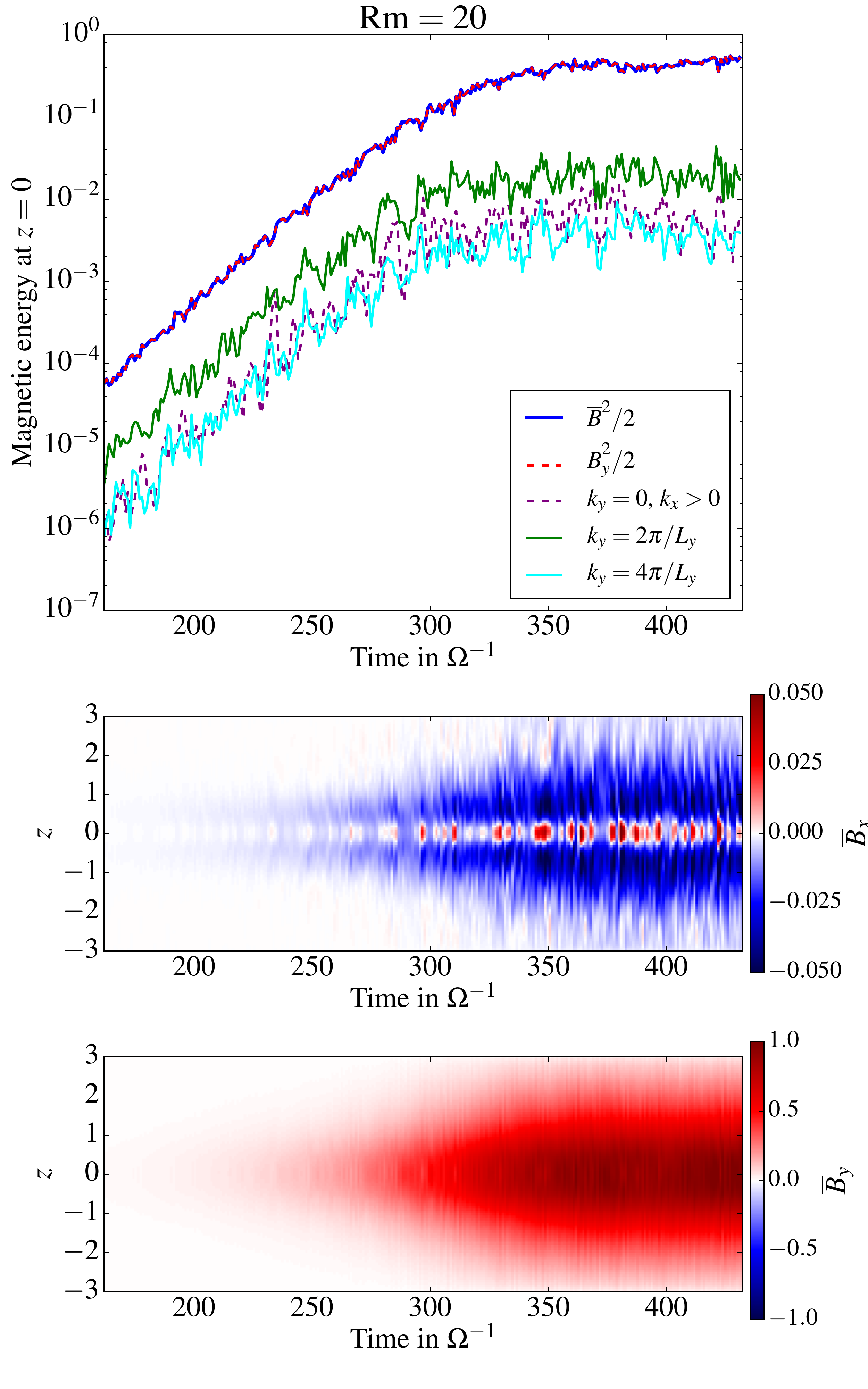}
\includegraphics[width=0.497\textwidth]{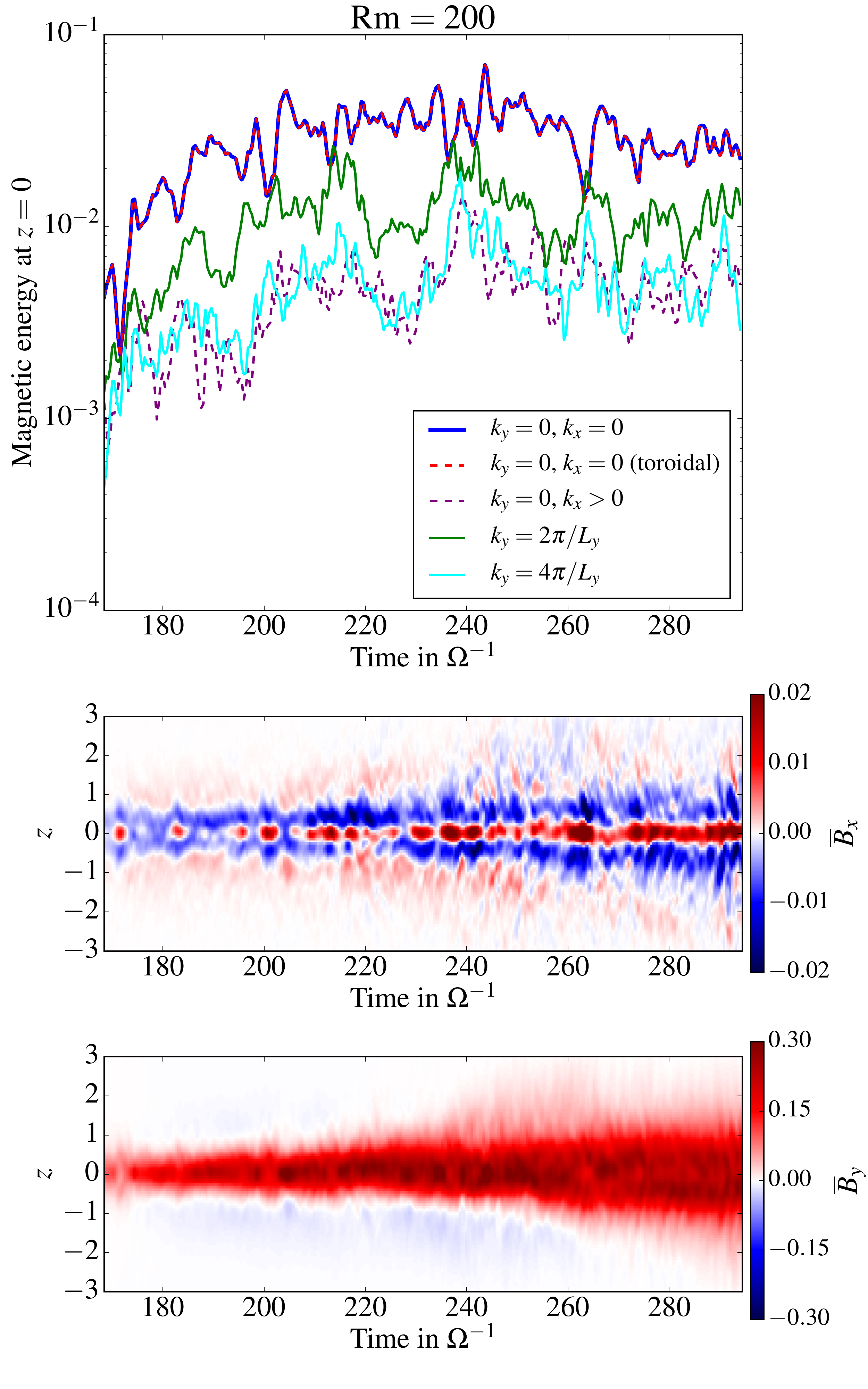}
 \caption{Top: time evolution of magnetic energy contained within the midplane ($z=0$), projected onto different Fourier components, for $\tau_c=20\,\Omega^{-1}$. Blue/solid lines correspond to the total magnetic energy and red/dashed lines to the total toroidal field $B_y$. Purple/dashed lines represent the magnetic energy projected onto the axisymmetric modes ($k_x\neq0$,  $k_y=0$). Green and cyan plain lines correspond to the energy of non-axisymmetric components, respectively $k_y=2\pi/L_y$ and $k_y=4\pi/L_y$.  Lower panels: space-time $(t,z)$ diagram of the mean $\overline{B}_x$ and $\overline{B}_y$. The  left panels are for $\text{Rm}=20$ and the right panels are for  $\text{Rm}=200$.}
\label{fig_fftmode}
 \end{figure*} 
 
 \begin{figure}
\centering
\includegraphics[width=\columnwidth]{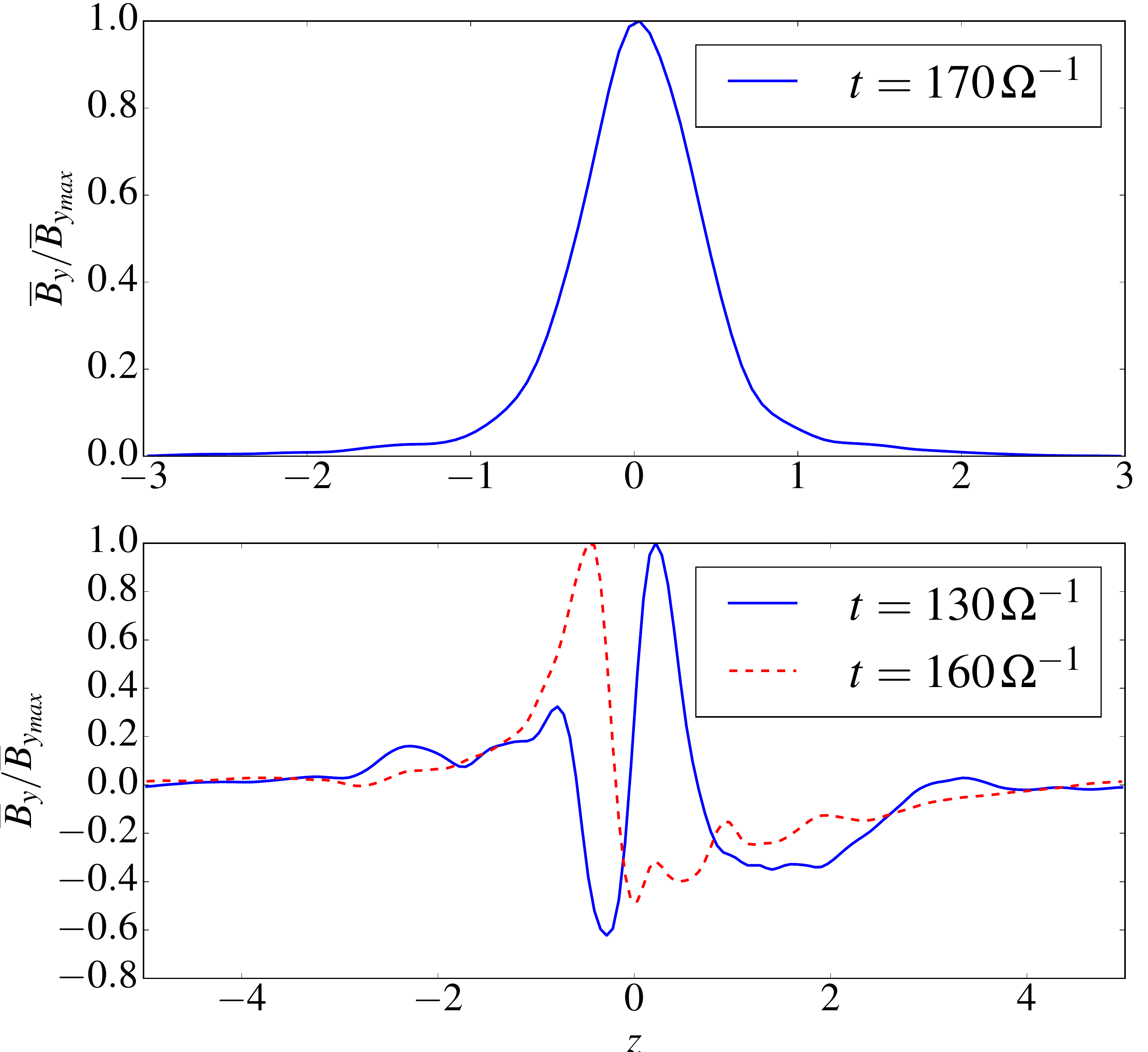}
 \caption{Vertical profile of $\overline{B}_y$ for $\text{Rm}=200$, with open boundaries (top) and periodic boundaries (bottom). In the latter case, two different times are represented to show the flipping of $\overline{B}_y$.}
\label{fig_byprofiles}
 \end{figure} 
  \begin{figure}
\centering
\includegraphics[width=\columnwidth]{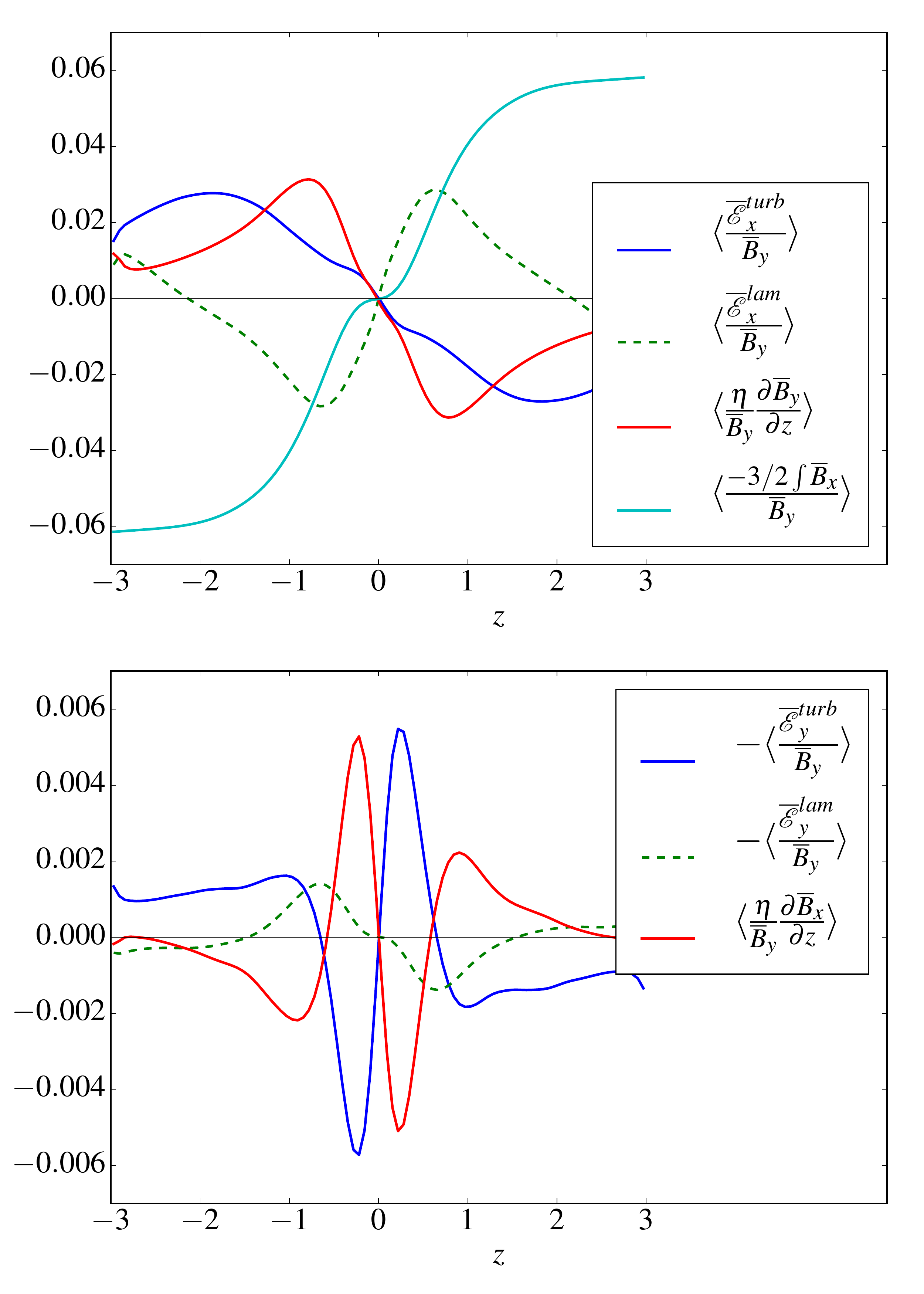}
 \caption{Vertical profiles of the electromotive forces, ohmic current
   and linear stretching term (omega effect) involved in the evolution
   of $\overline{B}_y$ (top) and $\overline{B}_x$  (bottom),  for
   $\text{Rm}=20$,  $\tau_c=20 \,\Omega^{-1}$ and open boundaries.  We
   remind the reader that $\mathcal{\overline{E}}^{lam}(z)= \overline{\mathbf{u}} \times \overline{\mathbf{B}}$ and $\mathcal{\overline{E}}^{turb}(z)=\overline{\mathbf{\tilde{u}} \times \mathbf{\tilde{b}}}$. All terms are normalized to the midplane $\overline{B}_y$ and averaged over $x$, $y$ and time, during the kinematic phase. }
\label{fig_emfs}
 \end{figure} 
   \begin{figure}
\centering
\includegraphics[width=\columnwidth]{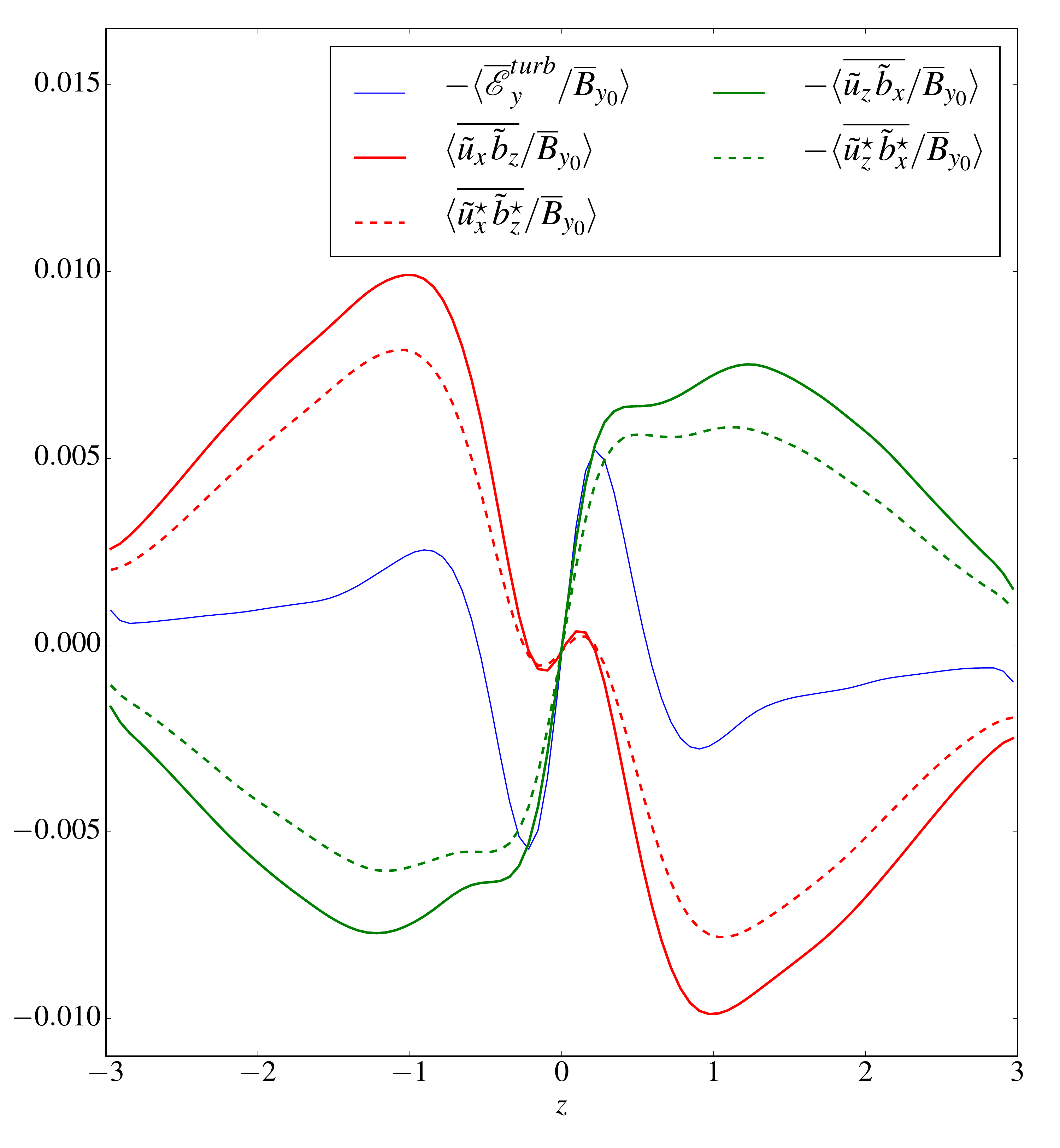}
 \caption{Vertical profiles of the toroidal EMF $\mathcal{\overline{E}}_y$ (blue line) and contribution of horizontal (red lines) and vertical (green lines) motions to $\mathcal{\overline{E}}_y$. The dashed lines are the EMF calculated by filtering out small-scale non-axisymmetric perturbations in Fourier space and retaining only the modes with $k_y\leq 2\pi/L_y$. All terms are normalized to the midplane $\overline{B}_y$ and averaged over $x$, $y$ and time, during the kinematic phase.}
\label{fig_emf_filter}
 \end{figure}

\subsection{Dependence on boundary conditions}
\label{boudary}

In the solar context, boundary
conditions strongly influence the  excitation and saturation of
convective dynamos \citep[e. g.][]{choudhuri84,jouve08,kapyla10}. We
now check if the GI dynamo exhibits a similar sensitivity. 
We compared three different vertical boundaries:
vertical field (open boundary), perfect conductor (closed boundary),
and periodic. 
Note that the perfect conductor has $u_z=0$, $B_z=0$ and
$dB_x/dz=dB_y/dz=0$. In theory, it prevents any EMF from being
generated at the boundary or outside the domain, and conserves
magnetic helicity in the box. However, in practice, given that the
EMF is reconstructed at the cell interfaces while boundaries are
given at cells centres,  the code still generates a non-vanishing EMF
at the boundary that can lead to a mean $\langle B_x \rangle$  and
$\langle B_y \rangle$.  When using such condition, we found similar
dynamo growth rates as in the open boundaries case,  but ended up with
higher magnetization in the saturated state.

Periodic boundary conditions are interesting since they ensure that no
EMF is produced at the boundaries and thus provide the best set-up to
test dynamo action as formally defined.  We superimpose in
Fig.~\ref{fig_growthrate} the growth rates obtained for three
different Rm (120, 200, 400) and also without explicit diffusion (diamond
markers). Although a dynamo appears,  the critical magnetic
Reynolds number is shifted to higher Rm, with $\text{Rm}_c\simeq
200$. Note that this effect appears to be independent of vertical box
size because we obtain a similar result with
$L_z=6H$ and $L_z=10H$.  As in the open boundary case, the field
grows quasi-exponentially, but growth rates and saturated magnetic
energy are lower. The fact that the dynamo depends
strongly on the nature of the boundary is actually not surprising. We
will show in Section \ref{emf_terms} that periodic boundaries prevent
flux from leaving the domain and even to be recycled from the upper
surface to the lower surface (and vice versa), which is somewhat unrealistic. One quantifiable effect is that the
large-scale magnetic field is forced to be antisymmetric about the midplane, which enhances ohmic diffusion.   

\section{Dynamo properties}
\label{sec_dynamo_scale}

In this section we turn our attention from the dynamo's broad-brush
features (growth rates, saturated amplitudes) and concentrate on some
of its specifics so as to better understand how it works. We focus 
primarily on the kinematic phase of its evolution.

\subsection{Large or small scale?}
\label{fft_modes}

To understand whether the GI-dynamo preferentially amplifies large or
small-scale fields, we show in Fig.~\ref{fig_fftmode} (top
panels)  the evolution of magnetic energy associated with different
Fourier modes when $\tau_c=20\Omega^{-1}$.  The calculation
is restricted to the midplane $z=0$ as it contains most of the
magnetic energy.  For both $\text{Rm}=20$ and $\text{Rm}=200$, the
magnetic field is clearly dominated by the mean component
$\overline{\mathbf{B}}$ ($k_x=0$, $k_y=0$), and especially its toroidal
projection $\overline{B}_y$.  The dynamo is thus predominantly large
scale, in the sense that magnetic fields grow on lengths much larger
than $H$.

The second important component is the fundamental non-axisymmetric
mode $k_y=2\pi/L_y$ (in green) which corresponds to a large scale
spiral magnetic structure (see section
\ref{sec_spirals}). Non-axisymmetric modes are essential to the dynamo
process according to the Cowling's anti-dynamo theorem, so we expect
them
to feature. In
particular, the interaction  of their magnetic and velocity components produce a
mean electromotive force that allows $\overline{\mathbf{B}}$ to be
constantly re-generated so as to compensate ohmic losses. Note that
axisymmetric modes with $k_y=0$ and  $k_x>0$ seem negligible in the magnetic energy
budget.

Cases $\text{Rm}=20$ and
$\text{Rm}=200$ are distinguished by their ratio of large to small scale magnetic
fields. In terms of energy, there is only a factor 4 difference between the mean
$\overline{B}_y$ and the non-axisymmetric harmonic $k_y=4 \pi/L_y$ for
$\text{Rm}=200$, while this ratio exceeds 100 when $\text{Rm}=20$. 
This we attribute to a drop in the mean
field $\overline{\mathbf{B}}$, the saturation level of the
non-axisymmetric mode being unchanged when Rm is increased.
So it would appear that the large-scale dynamo is weakened at large Rm,
and supplanted by a smaller-scale dynamics.  The origin of the small-scale magnetic fields is unclear at this
stage.  They might originate in a turbulent cascade of magnetic
energy, possibly from
local magnetic  instabilites, or from a small-scale
dynamo of Zel’dovich type that stretches and folds the field lines in
a random way. The mechanism behind the development of the small-scale
field is discussed in Section \ref{braided_smallscales}. We stress
that the MRI is not (or marginally) active in the very diffusive regime
considered in this paper, and is anyway likely to be strongly impeded
by GI  \citepalias[see][]{riols17c}.

\subsection{Vertical profile of mean magnetic field} 
\label{emf_terms}

In this subsection, we look in more detail at 
the properties of the large-scale dynamo during its \emph{kinematic} phase. 
The lower panels of Fig.~\ref{fig_fftmode} show the time-evolution of
the vertical profiles of the mean magnetic field $\overline{B}_x(z)$
(center) and $\overline{B}_y(z)$ (bottom) for $\tau_c=20\,\Omega^{-1}$
and open boundary conditions. For both $\text{Rm}=20$ and
$\text{Rm}=200$, the large-scale field is mainly toroidal
$\overline{B}_y/\overline{B}_x \simeq 20$ and remains confined within
the midplane region of the disc. This is reminiscent of simulations
without explicit resistivity  \citepalias{riols17c}. Unlike
the latter, however, the dynamo in resistive flow does not exhibit
time-reversals of $\overline{\mathbf{B}}$, or possibly these reversals
occur on timescales longer than $400\, \Omega^{-1}$. Figure
\ref{fig_fftmode} shows that both poloidal and toroidal components
have the same polarity in the midplane, but $\overline{B}_x$ changes
its sign at $z \simeq 0.2-0.3 H$. For $\text{Rm}=20$, the diagram
also reveals bursts of poloidal field confined to the midplane. 
These bursts are correlated with the
activity of vigorous density spiral waves. We note finally that for
large Rm, both the radial and toroidal mean field almost vanish in the
disc atmosphere, above one disc scaleheight.

Figure \ref{fig_byprofiles} compares the vertical profile of
$\overline{B}_y(z)$ produced from simulations with open and with
periodic boundary conditions,
for
$\text{Rm}=200$. Each profile is calculated at a given time of the
simulation during the kinematic phase. In the case of periodic boundaries,
the field is antisymmetric about the midplane, with a different polarity
in the upper and lower parts of the disc. This configuration ensures
that the mean $B_x$ and $B_y$ vanish in the box, but not over each
side of the disc. Therefore, the large-scale field in a periodic box
corresponds to the $n=1$ mode in $z$ (instead of $n=0$  for open
boundaries) which possesses a strong gradient at $z=0$. The vertical
scale associated with this gradient can be measured from
Fig.~\ref{fig_byprofiles} and is $\Delta z \simeq 0.2H$.  In
comparison the vertical scale of $\overline{B}_y(z)$ in open
boundaries simulations is 5 to 6 times larger.  As a consequence,
ohmic diffusion on $\overline{B}_y$ is enhanced by a factor of a few
tens, which should enhance the critical magnetic Reynolds number of
the dynamo. In addition,
horizontally averaged fields  $\overline{B}_x$ and $\overline{B}_y$
did not necessarily dominate the magnetic energy budget and
flipped
regularly in time, with a short period of 5-6 orbits (see
Fig.~\ref{fig_byprofiles}). 

\subsection{Electromagnetic forces}

We now investigate how the mean field is regenerated.  We focus
particularly on the simulation with open boundaries,
$\tau_c=20\,\Omega^{-1}$ and $\text{Rm}=20$. In Fig.~\ref{fig_emfs},
we plot the vertical profiles of the different terms in
Eq.~(\ref{eq_Bxmean}) and Eq.~(\ref{eq_Bymean}) governing the
evolution of $\mathbf{\overline{B}}$. This includes the laminar and
turbulent part of the mean electromotive force $\overline{\mathbf{u}
  \times \mathbf{B}}$, as well as the shear stretching (the ``omega effect'')
and ohmic diffusion  $\eta \mathbf{\overline{J}}$. All these terms are
averaged in time, during the kinematic phase (between $t=150$ and
$t=350\, \Omega^{-1}$), and over the horizontal plane.  Since their
evolution is exponential during the amplification phase, we normalize
them by the midplane $\overline{B}_{y_0}$ during the averaging
procedure. 

The top panel shows that the toroidal component
$\overline{B}_y>0$ is mainly amplified via the radial laminar EMF
(green dashed curve)
and the omega effect (cyan), since both have a positive gradient in
$z$. 
In contrast,
the radial turbulent EMF (blue) has a negative gradient in $z$, like ohmic
diffusion (red),  and therefore acts as a turbulent diffusion on the mean
field.  
Note that the radial laminar EMF $\simeq -\overline{u}_z
\overline{B}_y$ is  associated with compression of the disc due to a
mean inflow $v_z \simeq -0.04 \,\vert z \vert <0$. This inflow remains relatively
weak compared to the  r.m.s vertical velocity and is confined to the
midplane regions $\vert z \vert \lesssim 1H$. Further away in the disc
corona, $v_z$ is now positive and expels magnetic flux outward.
Although the EMF generated by the inflow near $z=0$ is comparable to
the stretching by the shear (omega effect), it is insufficient on its
own to sustain the toroidal field against ohmic and turbulent
diffusion. The omega effect is essential, and so we conclude that a poloidal
field is required for the large-scale dynamo to work.

The bottom panel shows the EMFs and ohmic diffusion profiles
associated with the radial field $\overline{B}_x$. Here the
laminar term (dashed green curve) is almost negligible compared to the
turbulent EMF  (blue). Moreover, the profiles are more complicated in
the bulk of the disk $|z|<H$. We point out that the toroidal and
turbulent EMF  $\mathcal{\overline{E}}_y$ has two local extrema of
opposite signs, one around $z \simeq 0.3 H$ and one around $z \simeq
1-1.5 H$. These extrema correspond to a change in the sign of the EMF
gradient, and then to a change of $\overline{B}_x$ polarity (see
Fig.~\ref{fig_fftmode} for comparison). Since both are correlated and
keep the same sign,  the turbulent EMF $\mathcal{\overline{E}}_y$  is
directly involved in the maintenance of the mean radial field against
ohmic diffusion.

\subsection{Nature of turbulent motions involved in $\mathcal{\overline{E}}^{turb}_y$}
\label{EMF_filter}

To proceed further in the analysis we need to understand the turbulent
EMF. In
Fig.~\ref{fig_emf_filter} we plot the part of the mean EMF
$-\mathcal{\overline{E}}^{turb}_y$ that issues from radial motions
$\overline{\tilde{u}_x \tilde{b}_z}$ (red curve) and the part that issues from
vertical motions $-\overline{\tilde{u}_z \tilde{b}_x}$ (green).
In the midplane ($\vert z \vert  \lesssim 0.3 H$) where $B_xB_y>0$,
the EMF is entirely due to vertical motions (such as upwellings and
rolls). 
In particular, this means
that the strong horizontal motions associated with spiral waves do not
play a direct role in the regeneration of the mean poloidal field
(though they play an indirect role, see Section
\ref{sec_spirals}). Higher up, in the disc corona, both radial and
vertical motions contribute to the EMF, with $\overline{\tilde{u}_x
  \tilde{b}_z}$ being slightly stronger and leading to the change of
$\overline{B}_x$ polarity.

To determine whether the mean field is generated by large-scale
motions (like spiral waves) or small-scale fluctuations, we calculate
the contribution of large-scale modes with $k_y\leq2\pi/L_y$ to the
EMF. In practice, this is done by computing the direct FFT of
$\tilde{u}_x$, $\tilde{u}_z$, $\tilde{b}_x$ and $\tilde{b}_y$,
filtering out all the Fourier modes that have $k_y>2\pi/L_y$, and
then going back to real space to calculate the products $\tilde{u}^\star_x
\tilde{b}^\star_z$ and $\tilde{u}^\star_z \tilde{b}^\star_x$, where
$^\star$ refers to a filtered component. This procedure removes the
correlation associated with small-scale non-axisymmetric turbulent
structures, such as helical waves generated by  parametric instability,
for instance \citep{riols17b}.  We superimpose in
Fig.~\ref{fig_emf_filter}  the filtered EMF (dashed lines) and show
that they represent approximately 80\% of the total EMF. This is
strong evidence that the mean field dynamo is supplied by motions
associated with large
scale spiral waves (with $k_y=2\pi/L_y$ and radial size  $\lambda
\simeq H$) rather than GI small-scale turbulence. 

In summary, we have demonstrated that the dynamo functions via (a)
the creation of $B_y$ from $B_x$ by the omega effect (mainly)
and (b) generation of $B_x$ by relatively large-scale, but turbulent, velocity fluctuations issuing from the spiral density waves.

\subsection{Mean field theory and its limitations}
 \begin{figure}
\centering
\includegraphics[width=1.04\columnwidth]{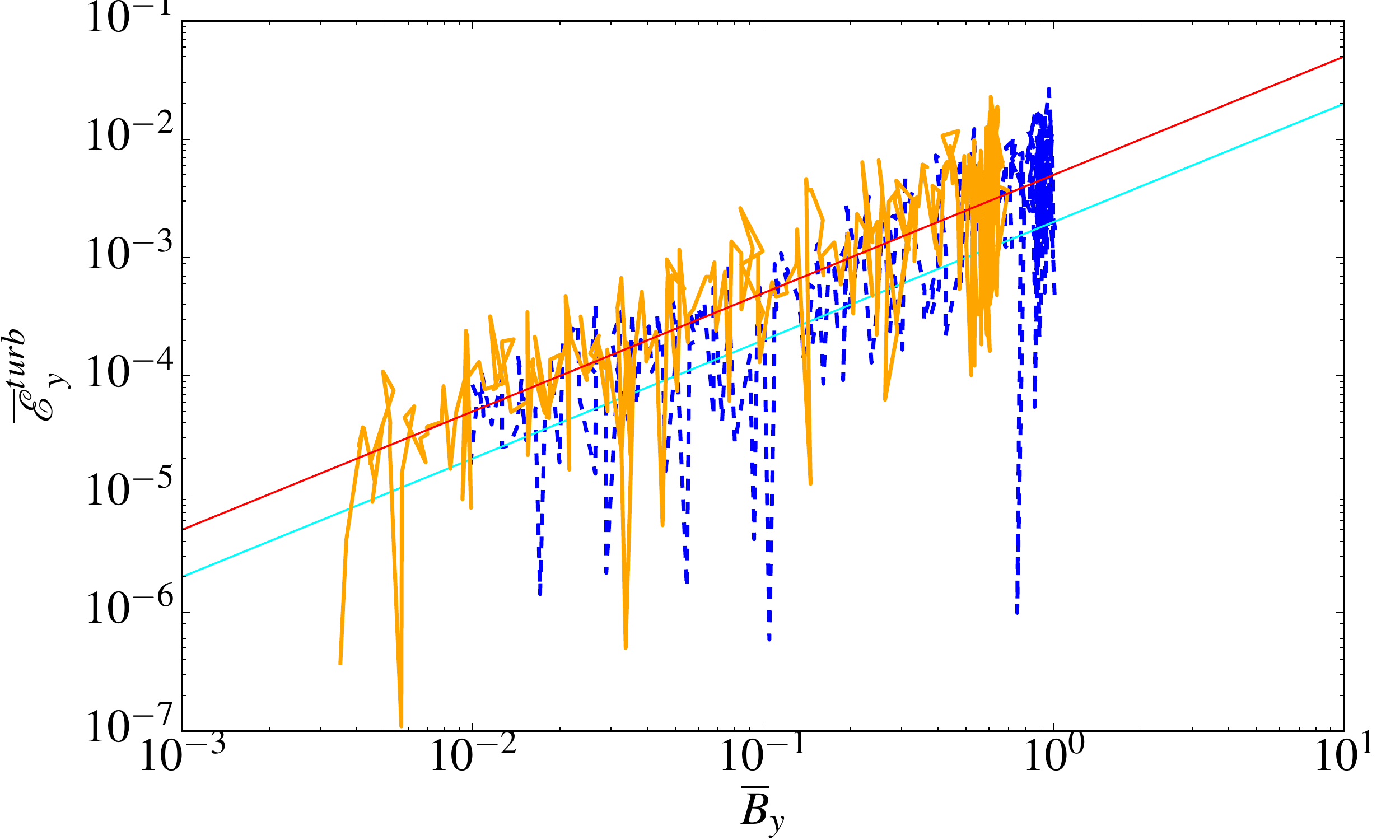}
 \caption{Phase portrait of the mean turbulent EMF $\vert \mathcal{\overline{E}}_y \vert $ versus the mean toroidal field  $\overline{B}_y$, during the whole simulation time, for $\tau_c=20\,\Omega^{-1}$ and $\text{Rm}=20$. The blue line is the EMF at $z=0.27 H$ (where it is maximum) and the orange line is at $z= H$ (where it is minimum). The dashed line corresponds to negative  $\mathcal{\overline{E}}_y $ while the plain line corresponds to positive $ \mathcal{\overline{E}}_y $. The red and cyan straight lines have respectively equations $y=0.005 x$ and $y=-0.002 x$. }
\label{fig_alpha}
 \end{figure} 

Figure \ref{fig_alpha} shows the evolution of the mean turbulent EMF
$\mathcal{\overline{E}}_y$ as a function of the mean toroidal field
$\overline{B}_y$, during the course of the simulation with $
\text{Rm}=20$, $\tau_c=20\,\Omega^{-1}$ and open boundaries.  Although
the EMF is highly fluctuating in time, on the average it scales
linearly with $\overline{B}_y$:
\begin{equation}
\label{eq_alpha}
\mathcal{\overline{E}}_y(z) \simeq \alpha_\text{dyn}(z) \overline{B}_y(z). 
\end{equation}
with  $\alpha_\text{dyn}$ of order -0.002 for $z\simeq 0.3 H$ and 0.005 for
$z\simeq H$. It is hence tempting to describe the dynamo process in
the framework of classical mean field theory
\citep{moffatt77,parker79,krause80,branden05,branden18}, according to which
the mean EMF is assumed to be a linear combination of the mean
$\mathbf{\overline{B}}$ and its vertical gradient. The coefficients of
the linear system encapsulate the properties of the turbulence and can
be derived using the so-called ``test field method". However, there are
several caveats. First, this theory only applies when
$\text{Rm} \ll 1$ or when there is a clear separation of scales between
the turbulent motions and the large-scale field. In our simulations, these
assumptions are far from satisfied, since the
fundamental non-axisymmetric mode associated with spiral arms
represents 80\% of the mean EMF (see \ref{EMF_filter}).  Second, we
showed that as the EMF changes its sign with altitude, so does
$\alpha_\text{dyn}$. Mean field theory predicts that $\alpha_\text{dyn}$ is
proportional to the mean helicity of the flow $\langle
\overline{\mathbf{u} \cdot \nabla \times \mathbf{u}} \rangle $, but
according to our simulations, the helicity keeps a constant sign on
each side of the disc and is therefore not correlated with
$\alpha_\text{dyn}$. One could probably find a more sophisticated model than
Eq.~\eqref{eq_alpha} , involving non-diagonal terms and dependences on
the mean field gradient, but this is not the object of the paper. 
 \begin{figure}
\centering
\includegraphics[width=0.8\columnwidth,trim=8cm 0cm 8.5cm 2cm, clip=true]{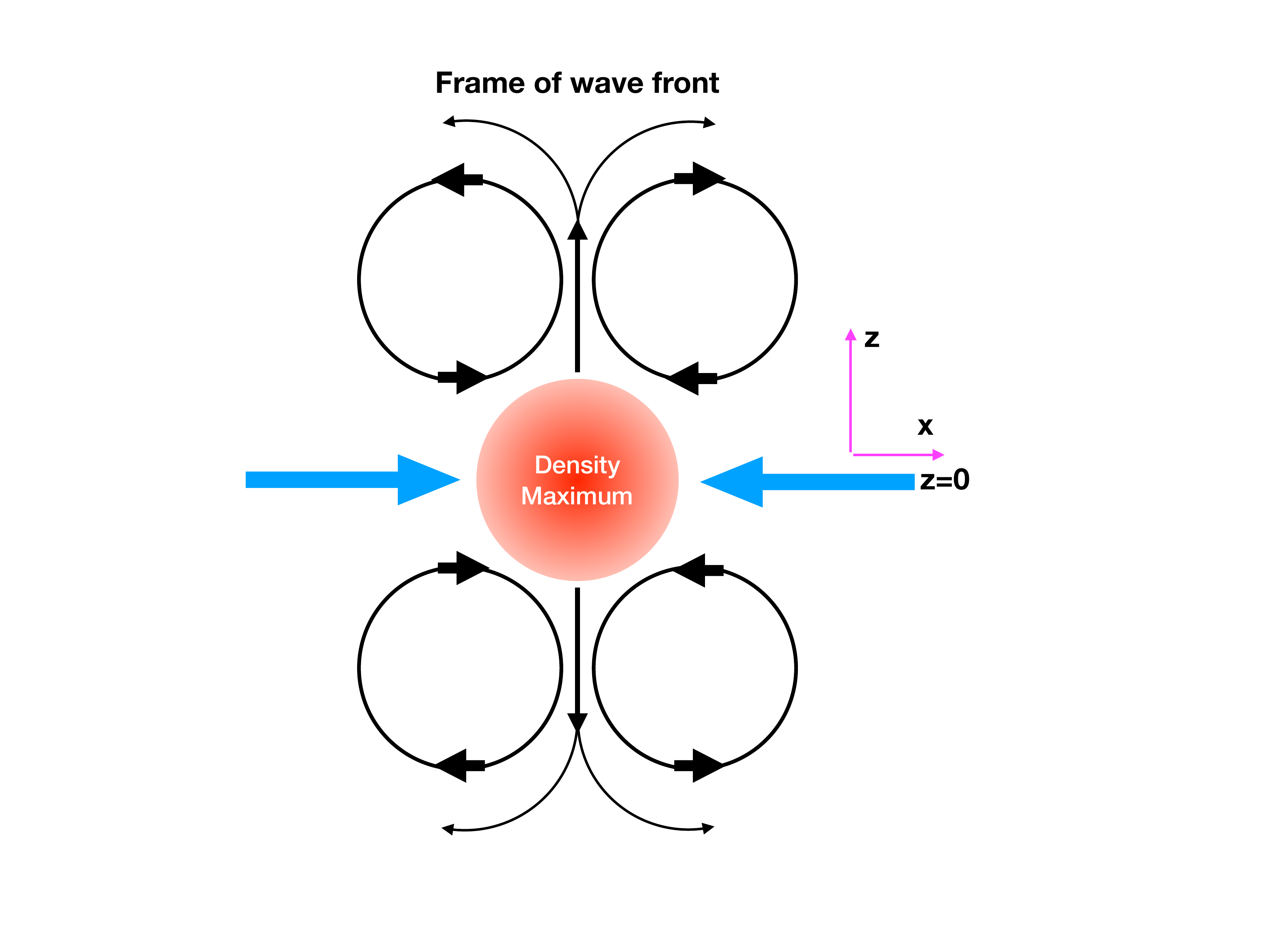}
 \caption{Sketch of the poloidal motions around a spiral density wave in the plane perpendicular to the wavefront.  The structure is invariant along the wave front, which has some pitch angle with the azimuthal direction $\mathbf{e}_y$. The blue arrows illustrate the horizontal 2D compressible motions, the black arrows the incompressible motions associated with the vertical rolls identified in \citetalias{riols18}. The structure is embedded in a background shear flow coming out of the plane (but with some pitch angle). Such rolls flow is reminiscent of the Roberts flow \citep{roberts72}.}
\label{fig_sketch_rolls}
 \end{figure} 
 \begin{figure*}
\centering
\includegraphics[width=\textwidth]{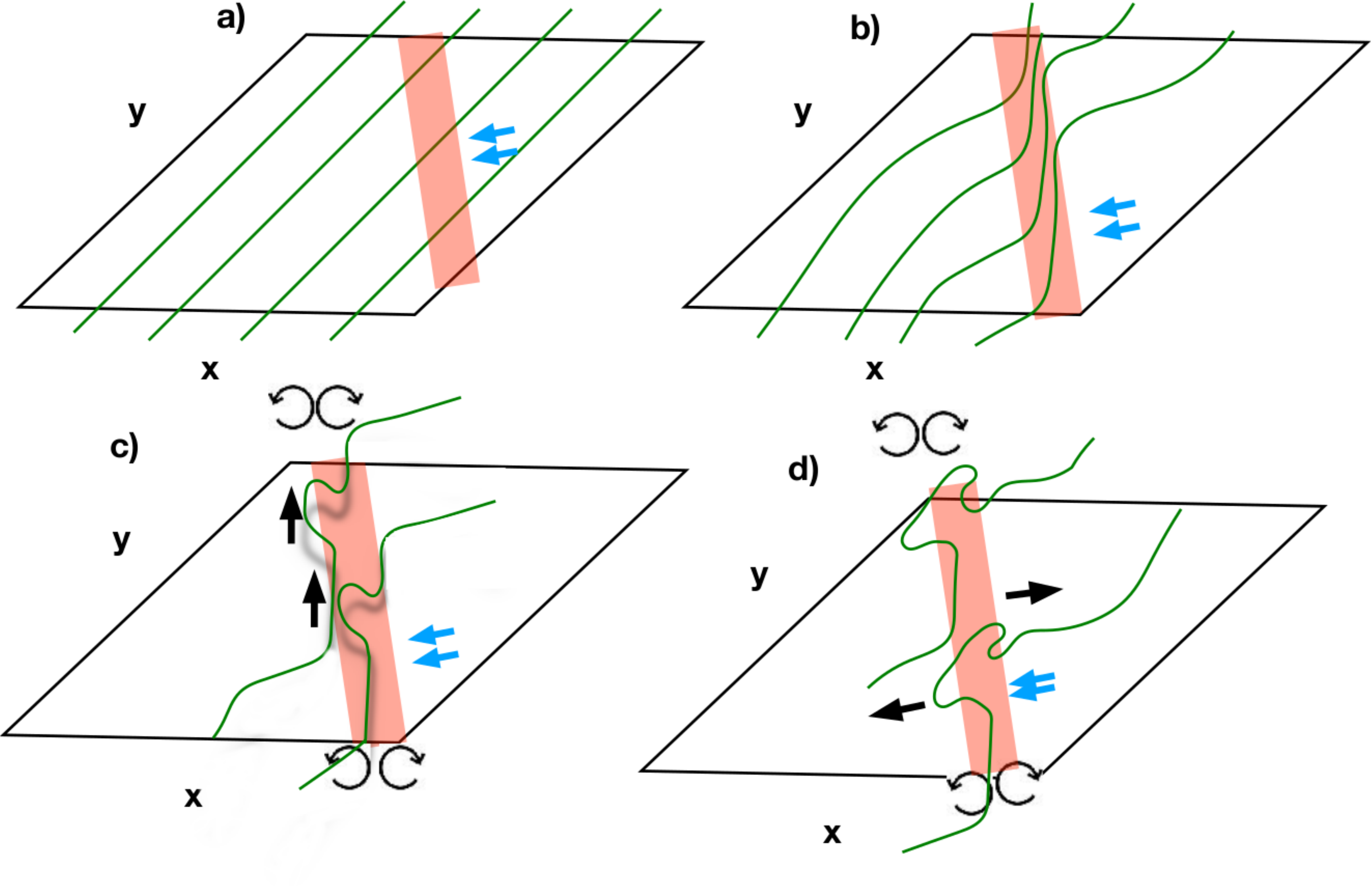}
 \caption{Sketch of the spiral dynamo mechanism. The magnetic field lines are represented in green while the red bar symbolizes a spiral wave. Black arrows denotes the direction in which magnetic field is stretched by the rolls. Each step is described in Section \ref{summary_dynamo}.}
\label{fig_sketch}
 \end{figure*}  

\section{Field amplification by density spiral waves: physical picture}
\label{sec_spirals}

We showed in the previous section that the GI efficiently
generates and sustains a large-scale magnetic field, especially in
the regime of low Rm $\lesssim 100$.  This dynamo survives at larger Rm, independently of the
vertical boundary condition, but produces weaker mean fields and seems to favour
smaller scales.  Our aim
in this section is to propose a physical picture of the dynamo action,
drawing
together the different results of the previous sections and
providing additional numerical evidence to illustrate the picture.

\subsection{Model of the large scale dynamo loop}
\label{summary_dynamo}

Our interpretation of the dynamo loop is very similar to
the ``$\alpha$-$\Omega$" mechanism proposed for
the solar dynamo cycle \citep{parker55}. As we showed earlier in Section \ref{emf_terms},
the toroidal field  is regenerated from poloidal field, through
an omega effect. This is straightforward and uncontroversial. 
The  poloidal field, however, issues from an ``alpha"
effect supported by large-scale GI spiral motions that stretch and fold the toroidal
field. Our task is then to elucidate how the
``stretching" and ``folding" motions occur, how they amplify
$\overline{B}_x$, and how these motions connect to the vertical EMF
profiles of Section \ref{EMF_filter}.  

Although the dynamo is a ``statistical'' effect comprising the work of
multiple spiral modes, it is better to start our investigation with a
single spiral wave and ask what it does to the magnetic field.  For
that purpose, we analyse the motion around a simple spiral
density wave.
Figure \ref{fig_sketch_rolls} sketches out its main dynamical
ingredients. The wave is composed of 2D horizontal
compressible motions in the midplane  (blue arrows)  and
incompressible vertical roll motions in the poloidal plane (black
arrows). These motions are drawn in the frame perpendicular to the
wave front, but are actually invariant along the direction of the
front. The pattern of four counter-rotating large-scale rolls has been
studied in detail by \citetalias{riols18}. These features have a
typical size $\simeq H$ and appear as a fundamental feature of linear
and nonlinear density waves in stratified discs. They can emerge from
a baroclinic effect associated with the thermal stratification of the
disc and are generally locally supersonic if the
wave is shocked. Such a circulation might also arise from hydraulic
jumps in very unstable flows \citep{boley06}. 
Lastly, there is the rotational shear flow coming out of the page,
though not perpendicularly.
Our pared-down sketch 
of the spiral wave flow is reminiscent of well-studied
and efficient kinematic dynamos such as those of Roberts and Ponamarenko type. It is then perhaps no surprise that we also find dynamo action
associated with the density waves.

Note that pure 2D horizontal motions, like those associated with wave
compression, are not able to sustain a mean field dynamo on their own.
It is true that locally they efficiently produce radial fields, but 
the mean $\overline{B}_x$
induced by these motions  (averaged over $x$ and $y$) necessarily
vanishes. Thus vertical flow structures, such as the counter-rotating
rolls, are essential: this can be seen immediately from the mean
toroidal EMF, a large fraction of which is supplied by either
vertical motions or vertical magnetic field (see section
\ref{EMF_filter}). 

Putting these ideas together, our  interpretation of the dynamo loop,
sketched in Fig.~\ref{fig_sketch},  is then the following: 
\begin{enumerate}[(a)]
\item Initially, the disc is threaded by a pure toroidal field with $B_y>0$. 

\item A spiral wave, tilted with respect to the azimuthal direction,
  passes through the disc. The field is then compressed horizontally
  by the spiral wave, along its pitch angle. On average, this process
  does not generate any mean radial field:  negative $b_x$ component
  is created within the spiral arms,  while a positive component is
  produced outside (in the inter-arm region).  See Section
  \ref{magnetic_patterns} for numerical evidence.

\item Where the field line crosses the centre of the spiral wave, it
  is pinched and lifted by the vertical velocity associated with the
  large-scale rolls. This process vertically redistributes the radial
  field and creates a mean $\overline{B}_x$ of opposite polarity in
  the midplane and in the corona (through the EMF term $\tilde{u}_z
  \tilde{b}_x$).  See Section \ref{vertical_rolls_mid}  for numerical evidence. A key point here is that the field in the wave isn't
perfectly aligned with the spiral front. Because of this non-alignment,
the field lines can be pulled apart by the two different rolls on each side of the wave.

\item As the rolls ``turn over" above the disk, they horizontally stretch and fold the field, producing a net $\overline{B}_x$ with
  opposite sign to $B_y$ in the corona (through the EMF term
  $\tilde{u}_x \tilde{b}_z$). See Section \ref{vertical_rolls_mid} for more details.

\item Finally, the mean radial field is stretched by the shear (omega effect)
  which produces a net toroidal field (not shown in
  Fig.~\ref{fig_sketch})
\end{enumerate}
Naturally, this theoretical
picture remains to be demonstrated or at least validated by numerical
simulations.  The next sections will focus especially on steps (b), (c) and
(d). 

\subsection{Numerical evidence of the spiral wave dynamo loop}
{To check that our dynamo model can be applied to GI turbulence,  we
analyse the magnetic field topology around spiral waves in our
GI simulations. All the following plots correspond to times when magnetic
energy grows linearly (i.e.\ in the kinematic phase). }

\subsubsection{Horizontal compression and magnetic spiral patterns: step (b)}
\label{3dgeometry}
\label{magnetic_patterns}
\begin{figure*}
\centering
\includegraphics[width=\textwidth]{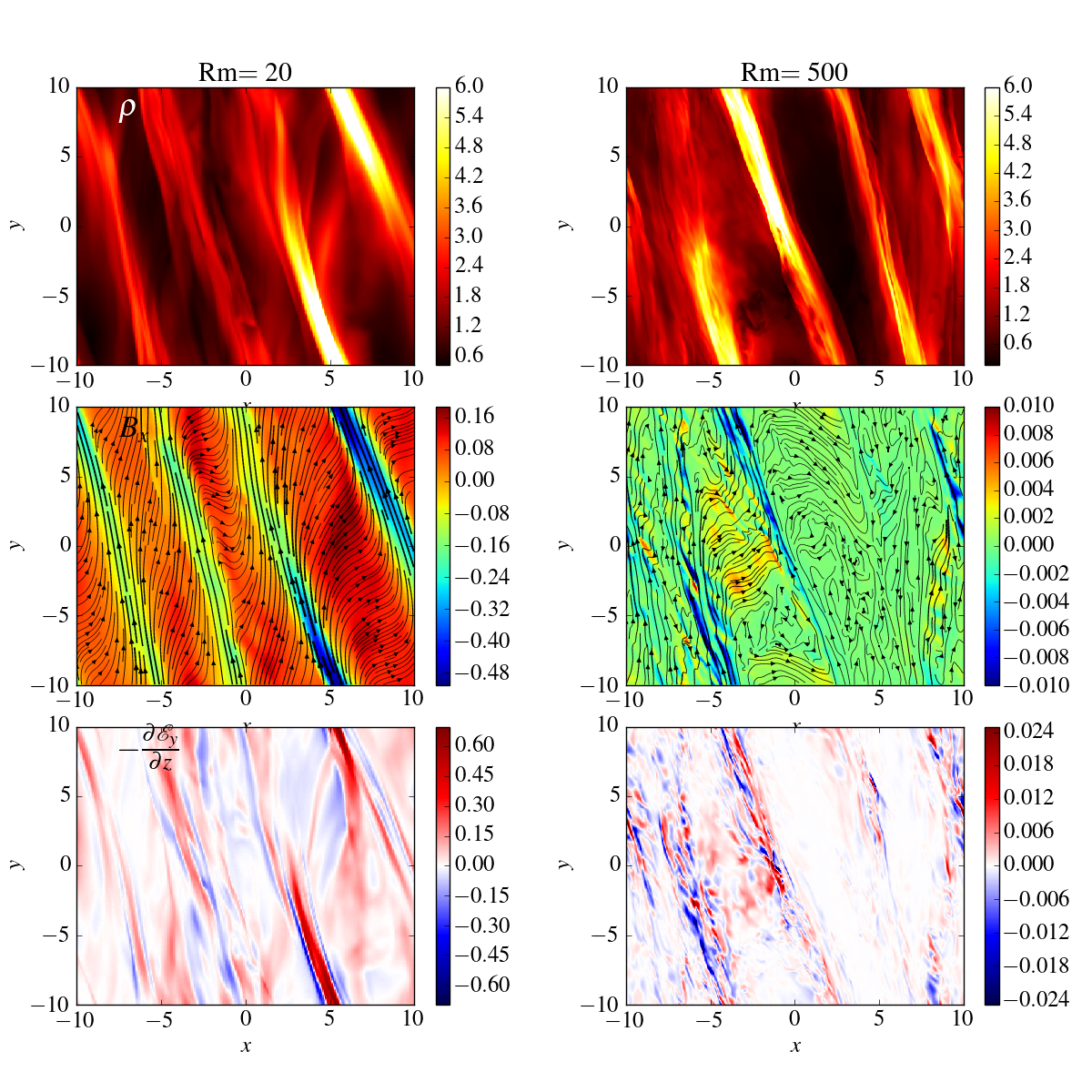}
 \caption{Density and magnetic field structures within the midplane ($z=0$), for $\text{Rm}=20$ (left panels) and  $\text{Rm}=500$ (right panels). Snapshots are taken at a random time of the simulation during the linear phase. From top to bottom,  $\rho$,  $B_x$ and the vertical derivative of the toroidal EMF $-\mathcal{E}_y$. In the center panels, the arrows are the horizontal projections of magnetic field lines. }
\label{fig_spiral}
 \end{figure*}

{
We first study the effect of spiral waves'  \emph{horizontal motion} on
magnetic field in simulations with $\tau_c=20\,\Omega^{-1}$. We
compare in Fig.~\ref{fig_spiral} the density structures in the
midplane (top panels) with the $B_x$ structures (center panels), for
both $\text{Rm}=20$ and $\text{Rm}=500$.   In both cases,  the radial magnetic field takes the shape of the density
spiral waves and concentrates within these structures.  In
particular, $B_x$ is always negative (as opposed to $B_y>0$) within
the density maxima, and positive outside the spirals. To complete the
analysis,  we superimpose the horizontal projection of magnetic
fields lines over the $B_x$ maps.  When $\text{Rm}=20$,
 the magnetic field is relatively strong and almost
aligned with the direction of the wave front inside the arms. Each field line, however, crosses the spiral wave  since it has to connect with the inter-arm field at some point (therefore, it is not perfectly aligned
with the front). }

{
 In the inter-arm
regions, the field is weaker and has a component perpendicular
to the fronts. This configuration is similar to that
illustrated in Fig.~\ref{fig_sketch} b) and results from the squeezing
of initially toroidal field by compressible motions within the waves. Because the  wave has a positive pitch angle 
\footnote{In turbulent shearing box simulations, the pitch angle is
generally around $5- 15 \degree$ and corresponds to the angle
for which most of the power from the shear is transferred to the
non-axisymmetric waves. Its relatively small value remains however
unclear and might depend on the  excitation process, wave phase
velocity and gas diffusivity.}, the field inside is tilted in the same
direction which leads to a negative $B_x$  (assuming an initially
positive toroidal field). In the simulation with $\text{Rm}=500$,
which has been run at double resolution, the field appears
fragmented and more filamentary.  However it
is again oriented along the spiral front with negative $B_x$ inside
the density maximum.
}
\subsubsection{Vertical rolls and step (c) and (d)}
\label{vertical_rolls_mid}
\begin{figure*}
\centering
\includegraphics[width=\textwidth]{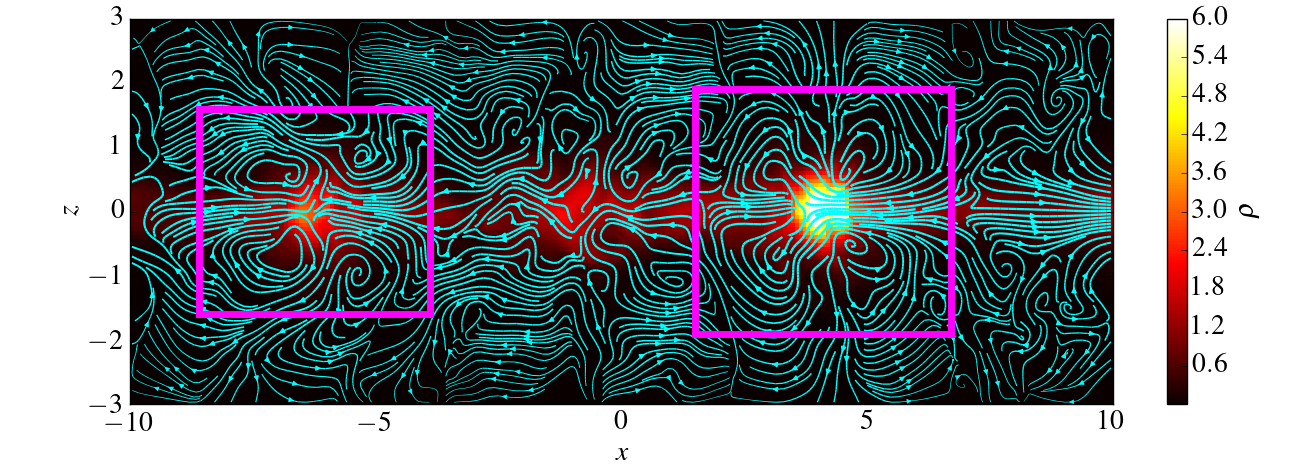}
 \caption{Top: density and streamlines  of the GI flow in the poloidal plane ($x,z$) for $\tau_c=20\,\Omega^{-1}$ and $\text{Rm}=20$ at a random time of the simulation during the kinematic phase. The thickness of the cyan lines account for the intensity of the poloidal mass flux. The pink rectangles are centred around the density maxima of two prominent spiral GI waves.} 
\label{fig_rolls}
 \end{figure*}

{
As suggested by our model (step c),  the vertical rolls surrounding each
density wave stretch the field lines vertically. 
 To show this effect numerically, we plot in Fig.~\ref{fig_rolls} (top)
the vertical structure of the gravito-turbulent flow in the poloidal
plane, at a random time. Around the most prominent density waves,
marked by a pink rectangle, we clearly identify a pair of
counter-rotating rolls, similar to those depicted in
Fig.~\ref{fig_sketch_rolls}.  }

\begin{figure*}
\centering
\includegraphics[width=\columnwidth]{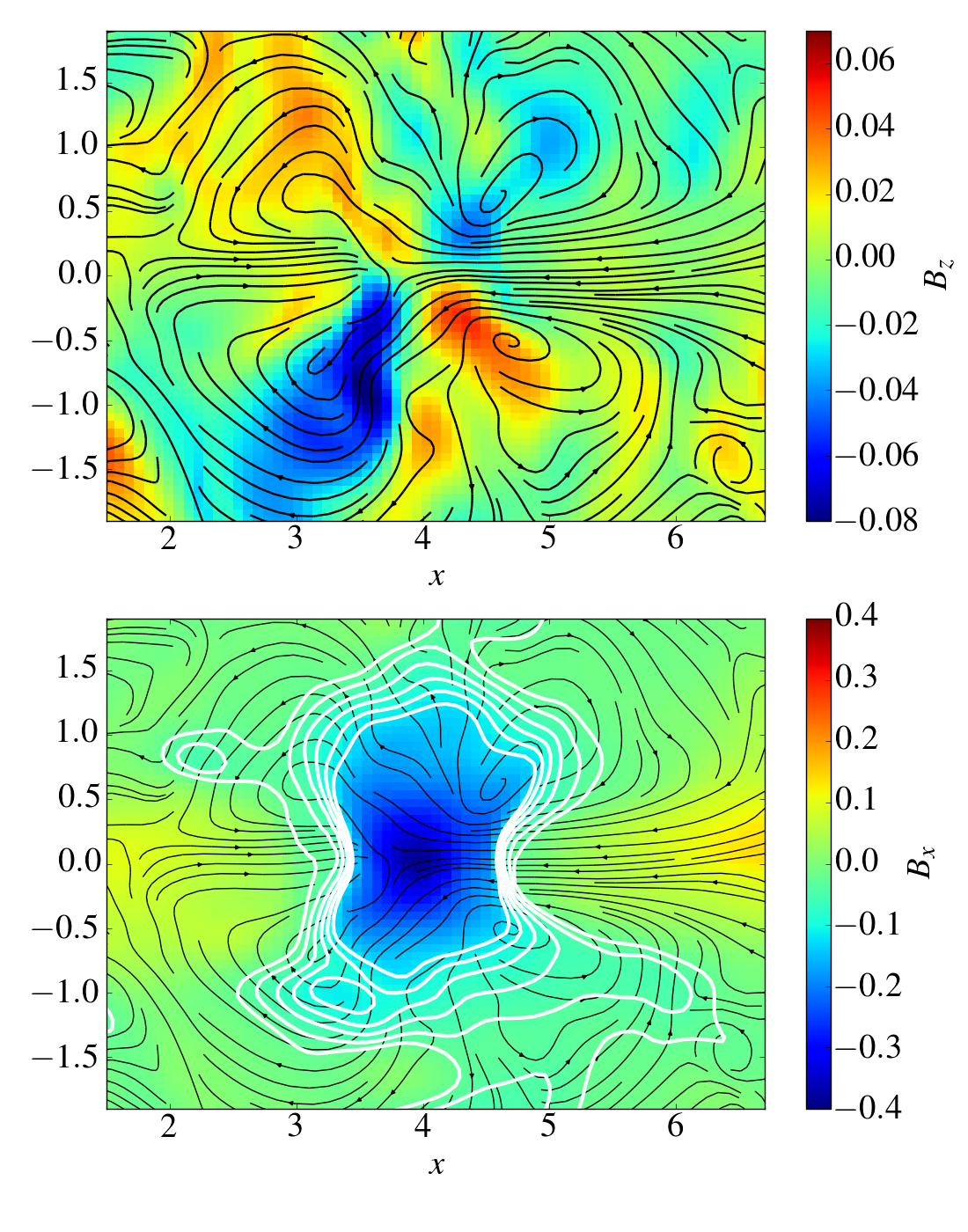}
\includegraphics[width=\columnwidth]{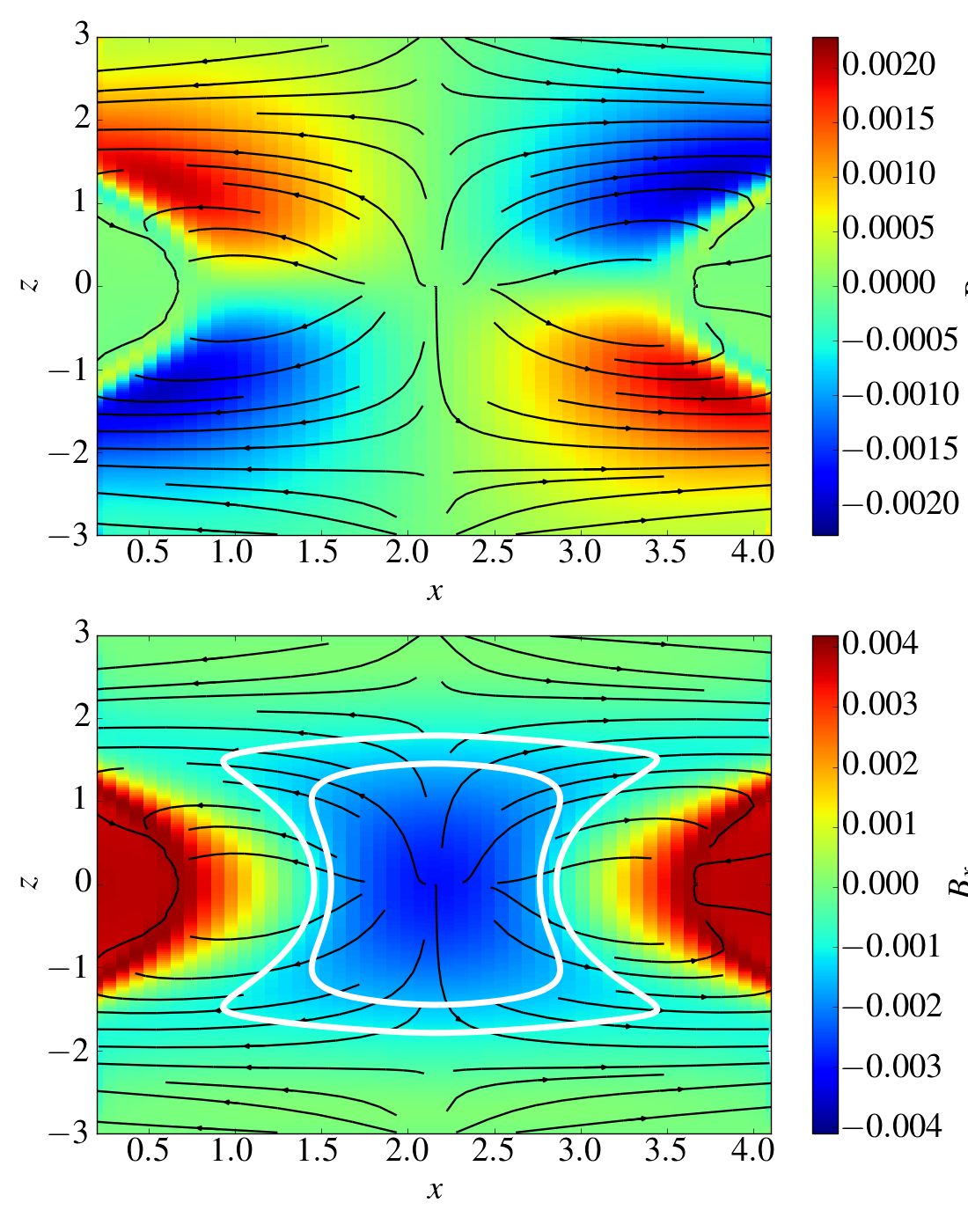}
 \caption{Magnetic field components $B_z$ (top) and $B_x$ (bottom) projected into the poloidal plane. The black lines are the poloidal streamlines and the white thick lines in the bottom panels are the contours of $B_x$ (showing the "mushroom" structure due to radial stretching). The left panels are zoomed around the second spiral wave of Fig.~\ref{fig_rolls} while the right panels are derived from a simulation of a simple spiral wave in polytropic atmosphere. In both cases, the magnetic configuration is invariant along the wave front.}
\label{fig_shwave_bxbz}
 \end{figure*}

{
The left panels of  Fig.~\ref{fig_shwave_bxbz} show $B_z$ (top) and $B_x$ (bottom)  around one of the rolls. For $z>0$,  as the field passes through the spiral wave (from ``left" to ``right"), it rises upward and produces  a positive $\tilde{b}_z$ component (shown in red). Once it reaches
the centre of the spiral, field lines move back toward the midplane $z=0$ and
finally connect with the inter-arm field, leading to a negative
$\tilde{b}_z$ (shown in blue) on the other side of the wavefront.  Counterintuitively, the magnetic field is not wrapped around the
centre of each roll, but is simply pinched at the centre of the wave,
where the vertical velocity of the roll is maximum, exactly as depicted in Fig.~\ref{fig_sketch} (c).  }

{
Ultimately, the field that rises up to $z\simeq H$ is
stretched horizontally by the counter-rotating motions of each roll and produces a
radial field component in the corona (step d). The stretching leads to
a ``mushroom-like" structure in the contours of $B_x$, reflected on either side of the midplane. These contours are plotted in white in the bottom left panel of Fig.~\ref{fig_shwave_bxbz}. } \\

{
In order to illustrate this process as cleanly as possible we simulate a simple laminar nonlinear
shearing density wave initially embedded in a uniform toroidal field
$B_{y_0}=0.02$ with $\text{Rm}=20$. The setup is the same
as in  \citetalias{riols18}: the wave is forced not by the GI but by a potential of
the form $\Phi_{\text{ext}}=A \,\cos\left[k_x(t)x+mk_{y_0} y\right]$
where $A$ is the amplitude of the forcing and
$k_x(t),k_{y_0}=2\pi/L_y$ are the radial and azimuthal wavenumbers.   We fix the amplitude $A=0.6$ in order to obtain a nonlinear shock wave. The initial pressure profile is polytopic
$P=K\rho^{1+1/s}$ with index $s=20$ (to mimic the entropy gradient
found in our GI simulations, see  \citetalias{riols18}). The spiral wave structure and its magnetic topology are shown  in the right panels of Fig.\ref{fig_shwave_bxbz}. The flow exhibits a roll structure, and this generates a cleaner version of the magnetic topology seen in the turbulent case (compare with the left panels). Again the stretching of the field gives rise to a mushroom structure, clearly visible in the bottom right panel of Fig.\ref{fig_shwave_bxbz}. }

\subsubsection{Spiral EMFs and generation of mean $B_x$}
{
To further develop the discussion in Section \ref{EMF_filter} and Fig.~\ref{fig_emf_filter}, we examine in detail the connection between the velocities and the EMF profiles. First, the bottom panels of Fig.~\ref{fig_spiral} show the derivative of the EMF
$-\mathcal{E}_y$ in $z$, projected onto the midplane. For
$\text{Rm}=20$, this quantity is concentrated within the density
spiral waves and has a positive feedback on the radial field.
Averaged over the box, this leads to a mean 
$\overline{B}_x>0$, consistent with Fig.~\ref{fig_fftmode}. 
In Section \ref{EMF_filter},  we demonstrated that such an EMF  is dominated by the $-\tilde{u}_z \tilde{b}_x$ component near the midplane. 
 Fig.~\ref{fig_shwave_bxbz} shows that at the location of the density maxima, where $\tilde{b}_x<0$, the vertical velocity of the rolls is always directed outward (positive $\tilde{u}_z$ if $z>0$ and negative
if $z<0$).  In the inter-arm regions, this turns to be the opposite configuration.  In summary, there is a 
clear correlation between $u_z$ and $b_x$ which directly generates a positive gradient of $-\mathcal{E}_y$. and then a positive $\overline{B}_x$ near the midplane region ($z<0.5 H$). Physically, this can be
interpreted as a vertical redistribution of radial magnetic flux. The
rolls expel the negative flux associated with the spiral outward, but
compress the positive flux  present in the inter-arm regions. This
redistribution happening globally around each spiral wave,  favours
the configuration depicted in Fig.~\ref{fig_fftmode}, with a
segregation between the midplane region (that builds up $
\overline{B}_x>0$) and the rest of the disc (that builds up $
\overline{B}_x<0$).  
}\\

{
This vertical redistribution is however incomplete, as we showed
that at higher altitudes, most of the radial field is produced through
horizontal motions, via the term $\tilde{u}_x \tilde{b}_z$ (see
Section \ref{EMF_filter}). Again there is strong evidence that the
rolls are involved in this process.  In fact, the maximum of
$\tilde{u}_x \tilde{b}_z$ (Fig.~\ref{fig_emf_filter}), located at
$z\simeq H$ corresponds exactly to where the radial velocity associated
with the rolls is maximum (this also corresponds roughly to their
vertical extent). Moreover, Fig.~\ref{fig_shwave_bxbz} shows that for $z>0.2-0.3 H$, the $\tilde{b}_z$ component is correlated with the rolls' horizontal motions. In regions where $u_x$ is negative, the magnetic field points
outward, and vice-versa. Such correlations lead to a negative EMF
$-\mathcal{E}_y=\tilde{u}_x\,\tilde{b}_z$ at $z>0$, resulting in a negative gradient for the EMF and therefore
a negative  $\overline{B}_x$, as expected from Fig.~\ref{fig_fftmode}
and \ref{fig_emf_filter}. }

\subsection{Emergence of small-scale braided structures in the large Rm regime}
\label{braided_smallscales}

We showed in Sections \ref{sec_threshold} and \ref{fft_modes} that the
mean field (or large-scale) dynamo is severely curtailed when
Rm increases. Instead, small-scale structures seem to emerge within the spiral waves while the mean field is diminished. 
We now explore how the flow develops the 
small-scale magnetic structures and why the mean-field dynamo is
optimal  at large resistivity.

A quick look at the lower panels of Fig.~\ref{fig_spiral} show
that the toroidal EMF for $\text{Rm}=500$ is still contained within
the spiral wave but its coherence on longer scales is lost. Parasitic small-scale modes interact with the large-scale magnetic structures depicted in
Fig.~\ref{fig_shwave_bxbz} and subsequently reduce the efficiency of the mean
field dynamo.  

\begin{figure*}
\centering
\includegraphics[width=0.95\textwidth,trim={4cm 1cm 4cm 1cm},clip]{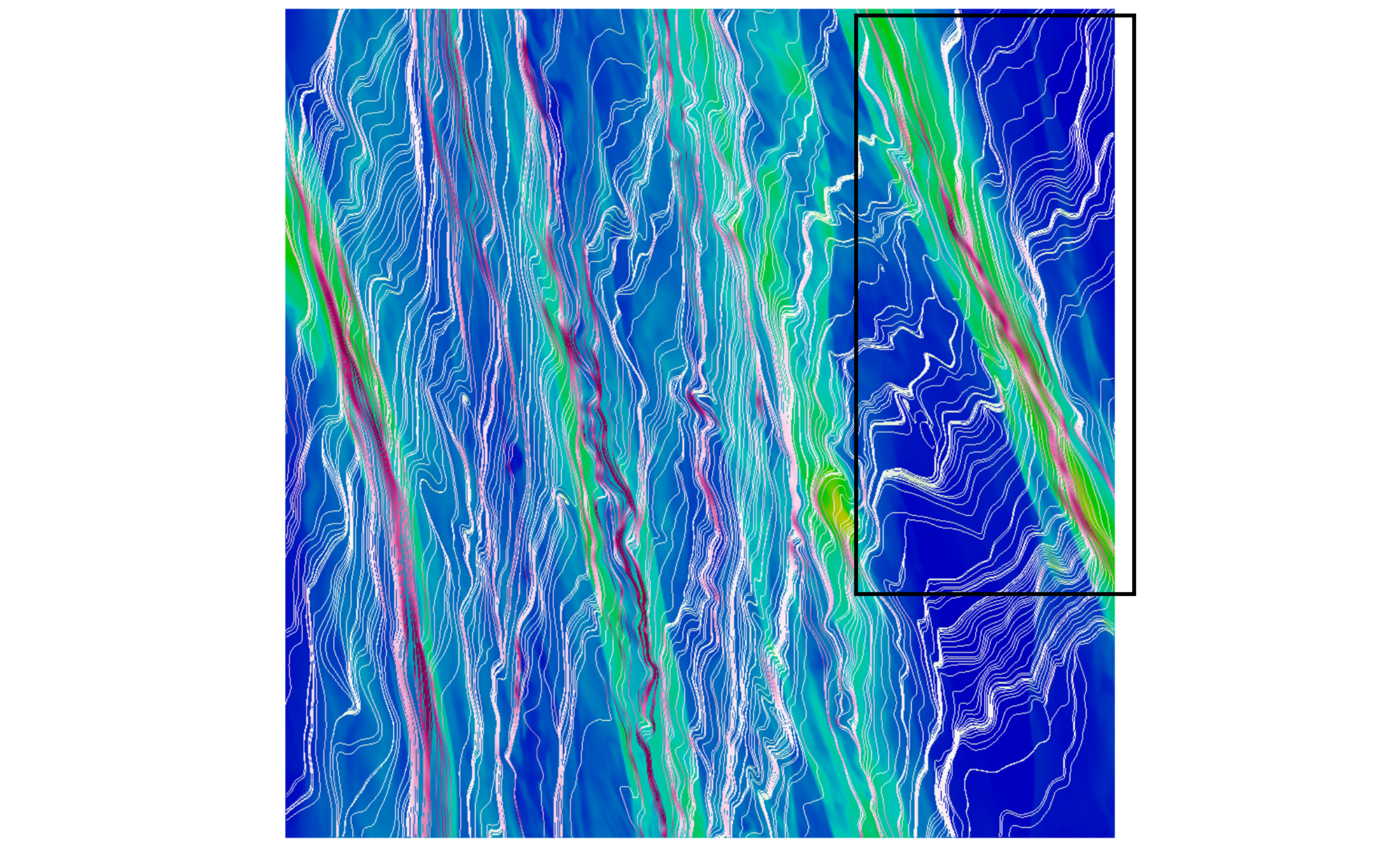}
\includegraphics[width=0.52\textwidth]{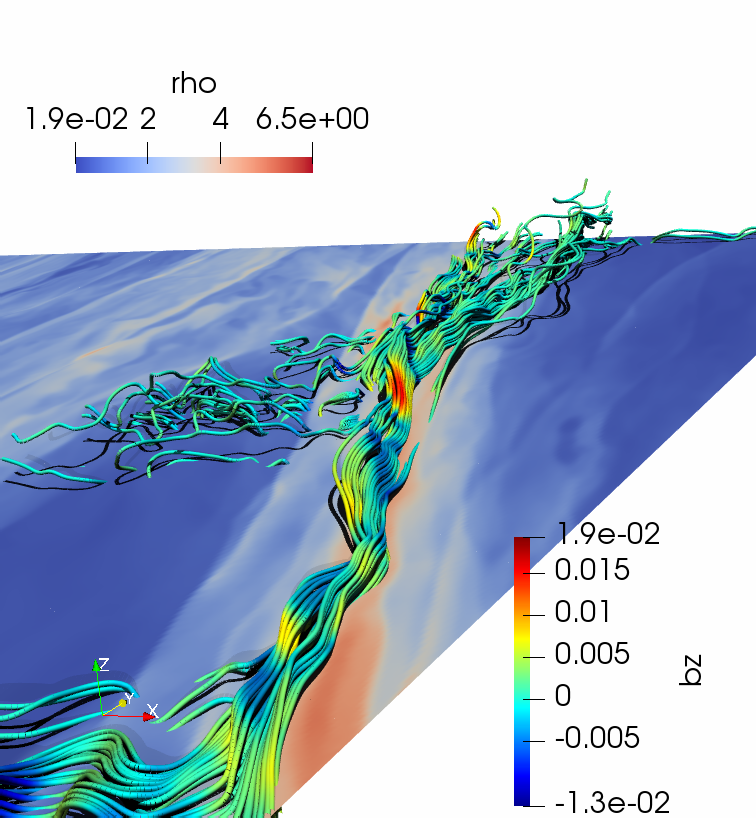}
\scalebox{-1}[1]{\includegraphics[width=0.47\textwidth, angle =90 ]{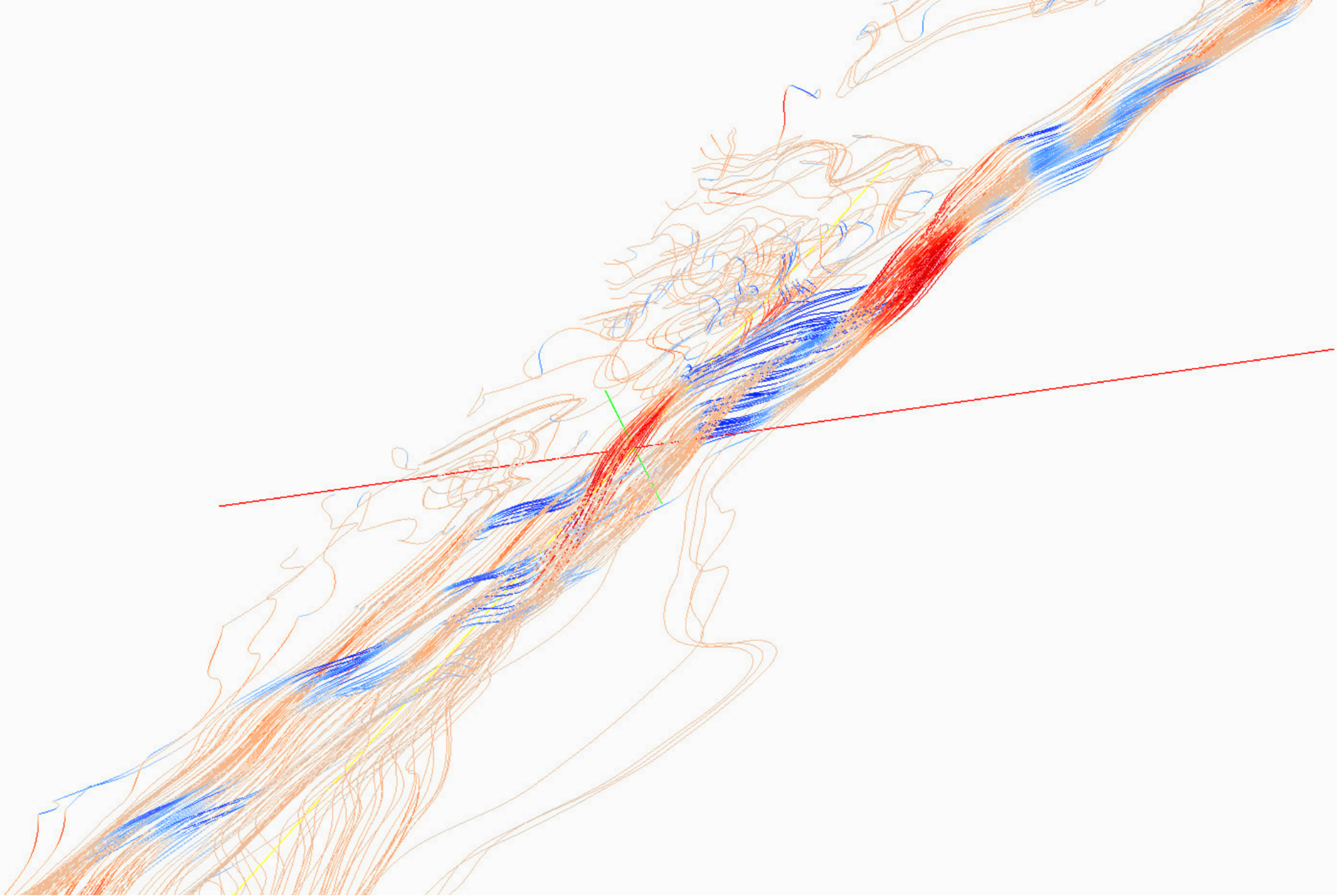}}
 \caption{Top: magnetic field lines in the midplane for $\tau_c=20\,\Omega^{-1}$ and $\text{Rm}=500$. The blue/green colours represents the gas density (green patterns are spiral waves). The white/pink lines are the magnetic field lines projected into the midplane. Pink color corresponds to high magnetic amplitude, white to low amplitude. The magnetic structure around the most prominent spiral wave  (within the black rectangle)  is shown in 3D in the lower panels from different angles.  The colour of the field lines denotes the amplitude of the vertical field $B_z$. The colours in the plane of the left panel indicate the midplane density.}
\label{fig_braided}
 \end{figure*}

We
show in Fig.~\ref{fig_braided} (bottom) a 3D rendering of the field line
topology around a given spiral wave, chosen randomly in our GI
turbulent simulation with $\text{Rm}=500$. To aid interpretation,
we show in the top panel  the original snapshot containing the full
domain with a 2D projection of the field.  There is clear a evidence
that field lines are rolled and highly twisted within the spiral wave,
forming a braided magnetic structure along the wave front.  Note that
a single structure is formed and settles at the intersection of the
four counter-rotating rolls, near the density maximum. The twisting
probably results from the combined effect of the differential rotation
and the helical flow associated with the spiral rolls. Such complex
and intricate structures do not appear in resistive runs, probably
because they relax rapidly through diffusion.

Twisted magnetic fields are known to be important in in the
context of solar flares and coronal heating, because they are subject
to kink instabilities when the
twist exceeds a critical value
\citep{dungey54,roberts56,hood81,bennet99}. The kink instability
mainly deforms and breaks magnetic tubes, and thus efficiently
produces turbulent fields and ultimately magnetic energy dissipation. 
If we define
$p$ to be the pitch of the twisted tube and $R_t$ its radius, the criterion
for instability is 
\begin{equation}
p R_t=\dfrac{\vert \nabla \times \mathbf{B} \vert}{\vert\mathbf{B}\vert } R_t  \gtrsim1,
\end{equation}
\citep{linton96}. In our simulations this condition is
marginally satisfied, which itself revealing. Though this instability criterion has
been derived with simple assumptions and well-defined equilibria
(which probably do not apply to GI flows), we believe that
braided magnetic fields fields within spiral waves are important in the
the generation of small-scale field and the breakdown of the mean
field dynamo at larger Rm. Stronger resistivity prevents magnetic field from being
twisted into a large number of loops and therefore keeps magnetic
energy from cascading to small scales.

\section{Discussion}
\label{sec_discussion}
 \subsection{Large-scale dynamo}
We showed that large-scale fields result from  an ``alpha" effect which involves vertical folding and horizontal
stretching by the rolls associated with spiral waves. 
The connection between the alpha effect and helicity has often been pointed out, and indeed it can be shown that the spiral  rolls are helical, since the wave
generates prograde motions where the roll is anti-clockwise and
retrograde motions where it is clockwise. As mentioned earlier, the helical
motion bears strong similarities with the Roberts flow \citep{roberts72}. But the  dynamo loop described above differs significantly from
dynamos supported by small-scale subsonic turbulent flow, such as
convection in the Sun. The motions here are supersonic, large scale
and highly localized in radius, which leads to a rather patchy and
inhomogeneous $B_y$, different from the initial uniform $B_y$. Fields
are inevitably filamentary and sparse in such GI flows.  Finally, we point out that 20\% of the mean field might also come from small-scale
turbulence, or inertial and helical waves triggered by a parametric
instability. 

\subsection{Boundary conditions}

One concern is that the dynamo behaviour differs when
using different boundary conditions (see  Section \ref{boudary}). In
particular, the strong and large-scale dynamo at low Rm is conditioned
by the existence of a net toroidal field throughout the box. However
we think that this issue is not critical for several reasons.  

First, and perhaps most obviously, the dynamo still works with
periodic boundaries. This suggests that magnetic field amplification mechanism is
not an artefact of the boundary.  In particular, the growth rates
obtained with open boundaries increase with decreasing $\tau_c$,
which is physically expected. The EMF at the boundaries in that case
is relatively small compared to the EMF near the midplane  (see
Fig.~\ref{fig_emfs}).  In a periodic box, we also checked that a 
large-scale field $\overline{\mathbf{B}}(z)$  can be generated, although it
exhibits a weaker amplitude and flips regularly, every 5 or 6 orbits.
Perhaps, one key question is to understand whether the dynamo in this
configuration is supported by the large-scale
$\overline{\mathbf{B}}(z)$  or by a small-scale  process of
Zeld’ovich type. The answer is not obvious, since the Rm regime
allowing the dynamo is marginally resolved by our simulations. In any
cases, the stretching and folding by the spiral rolls must be crucial
in the dynamo mechanism, whatever kind it is.  

Second,
boundary conditions are part of the physical problem, and open
boundaries are the most realistic of the options available to us.  
Periodic boundary conditions introduce
unnatural communication between the two surfaces of the discs which
can alter the nature of the large-scale field, as we see.
Moreover, periodic boundaries do not allow magnetic buoyancy to
evacuate excess magnetic field; its continued build-up may be unphysical.
They also force a particular geometry upon the field, which can
artificially enhance its dissipation. 
In real systems, the disc
exchanges field with an external ``corona'', though the details of how
this works is not easily determined.

\subsection{Application to protoplanetary discs}

The dynamo identified in this work is perhaps most relevant
for young and massive protoplanetary discs, subject to
GI. Indeed, except for the innermost regions ($\lesssim$ 1 AU), the
coupling between the gas and magnetic field in these
object is believed to be weak, and non-ideal MHD effects should be
important throughout, and certainly in the regions susceptible to gravitational
instability (typically beyond 20 AU). According to \citet{simon15},
using the standard MMSN disc model, the magnetic Reynolds number
in these regions varies between $10^3$ and $10^6$. However these
numbers must be considered gross upper limits for the 
denser and flatter discs susceptible to GI
\citep[see the case Elias 2-27 for instance,][]{perez16}.  The reason
is that FUV cannot penetrate very far into the disc when the surface
density is high.  We expect then  that some of the Rm probed by our
simulations are relevant.

In any case,  ohmic diffusion is not the most dominant
non-ideal effect at radii larger than 20 AU. Ambipolar diffusion is
believed to be the leading effect in these regions with typical
Reynolds (or Elsasser) numbers between 1 and 1000 \citep{simon15}. Although our
numerical set-up does not capture ambipolar diffusion,  it is not
excluded that its effect may be similar to ohmic diffusion in
the nonlinear phase of the dynamo, at
least if the magnetic field is not highly anisotropic.
We point out again that during the dynamo's kinematic phase, both the
Hall effect and ambipolar diffusion are subdominant to ohmic
resistivity, though this may change if there is a net magnetic field. 
Future
simulations are necessary to address how
ambipolar diffusion and the Hall effect influence the GI dynamo process, but also more generally
on GI turbulence.
For example, it is possible that
zonal flows enhanced by non-ideal effects \citep{lesur14,bai15}
significantly perturb the spiral waves motions or become even unstable
to non-axisymmetric perturbations \citep{vanon16}, leading to
different flow behaviour. 

It has been postulated that somewhat older disks may periodically undergo GI in their 
magnetically inactive dead-zones (located between roughly 1 and 10 AU)
during outbursts of FU Ori or EX Lupi type \citep{gammie96, armitage01, zhu10}. In affected regions the magnetic Reynolds number is low,
and a GI dynamo is a distinct possibility in the initial stages of an
outburst, 
perhaps supplying the seed field for the onset of MRI. The interplay
of the GI, GI dynamo, the MRI, and the Hall effect are expected to
present some intriguing dynamics. 

{
We finally give estimates of the field intensity in PP discs that
a GI-dynamo might be able to generate.   In the range of Reynolds number probed by our
simulations,  we found that the ratio of magnetic to thermal pressure $P$  lies in a range between 0.01 and 0.5 (depending on Rm).  We have $P = \rho c_s^2/\gamma  \simeq  \Sigma  H \Omega^2/(\sqrt{2\pi})$.  We consider a typical disc with surface density}
\begin{equation}
\Sigma \simeq 810\, \left(\frac{R}{\text{1 AU}}\right)^{-0.8}$ g/cm$^2
\end{equation}
and $
H/R \simeq 0.04 $
{
in agreement with observations of Elias 2-27  \citep{cieza17}, a massive disc susceptible to GI. By considering a disc with a central star mass equal to that of the Sun, we then hobtain}
\begin{equation}
B \simeq 0.4 - 2.5   \left(\frac{R}{\text{1 AU}}\right)^{-1.4}   \text{Gauss}
\end{equation}
{
At $R=30$ AU  we estimate the field strength to be around $B\simeq 3\, - 20\, m\text{G}$. 
We remind the reader that such fields are mainly toroidal, if generated through GI spiral waves. }

\subsection{Application to AGN}

The GI dynamo could be also excited
in the regions of AGN beyond 0.01 pc, where self-gravity is
thought to dominate \citep{menou01,goodman03,lodato07}.
The typical Rm in these objects at these radii is poorly constrained:
though the disk is very large, the ionisation fraction could dip to
low levels as the gas is too cool to support significant
thermal ionisation. One must then appeal to photoionisation, the efficiency
of which is uncertain at the midplane, and partly controlled by the morphology of the disk.
It is likely, however, that typical Rm values are
greater than those probed here. The same claims can be made with respect
to the ambipolar Reynolds number \citep{menou01}. Thus making a direct comparison with AGN is not straightforward.

The saturated magnetic energy of the dynamo never exceeds the thermal energy (cf.\ Fig. 4), 
but can nonetheless reach significant levels, with $\beta_t \sim 2$.
 Concurrently, the saturated Toomre Q does attain somewhat larger values (Fig. 5). 
We might then conclude that a saturated GI dynamo, when up and running, could help prevent 
a disk from fragmenting when it might otherwise. The thin AGN disk would then extend to 
larger radii than current hydrodynamical estimates predict, though it 
is unlikely that the dynamo's magnetic pressure on it own can completely solve the disk 
truncation problem \citep[see][]{goodman03}. Adding a vertical magnetic field of sufficient strength 
may be one possible solution here \citepalias{riols17c}. 
(Note that a completely analogous discussion can be had regarding 
giant planet formation in PP disks.) In any case, it is certainly likely that the GI and its
interaction with magnetic fields are foremost elements in the mysterious dynamics of the broad
line regions of AGN.

\subsection{Applications to galaxies}

Multiwavelength radio observations suggest that some spiral galaxies
(M81, M51 and possibly M33) are dominated by bisymmetric-spiral
magnetic fields. These magnetic spiral patterns are strongly
correlated with the large scale galactic spiral arms.  

To explain the correlation,  \citet{chiba90} and
\citet{hanasz91}  proposed a parametric swing excitation due to
resonances between weak magnetic wave oscillations and density spiral
waves. More sophisticated models of this resonance have been studied
\citep{moss97,rodhe99}, but most
 are based on a mean-field theory, in which the
interstellar turbulence, driven by supernovae, is
parametrized by an alpha effect \citep{elstner00,
  rudiger04,gressel08}. On the whole, the mean field theory provides
an acceptable match to the observations, though is not without its
challenges \citep[see, for example][]{nixon18,tabatabaei16}. 
Alternative theories, based on the vertical shear of
a primordial poloidal field, have been also proposed but remain
highly speculative \citep{nixon18}. Surprisingly, density spiral waves have rarely been
considered as capable of amplifying magnetic fields on their
own. Instead density waves
have been historically treated as a process that could shape large-scale fields
already generated. 
{Only recent MHD simulations of galactic discs have
   suggested some
 link between the dynamo and the spiral arms \citep{dobbs16,Khoperskov18}}.  Our
GI-simulations show that such waves can, in principle, act as a dynamo and produce
spiral magnetic patterns, similar to those observed in galaxies. 

Although there is a consensus that GI is active in galaxies, there is
however some uncertainty regarding the magnetic diffusivity. The
microscopic or Spitzer resistivity in galactic disc is about $10^8$
cm$^{2}$/s \citep{parker72}, which yields gigantic
magnetic Reynolds numbers. On the other hand, we could invoke
an anomalous resistivity due to small-scale
turbulence in order to explain dissipation of small
scale fluctuations in the ISM \citep{hanasz08}. The common value used
is  $\eta = 10^{26}$ cm$^{2}$/s  (300 pc$^2$
Myr$^{-1}$), which presents magnetic Reynolds number between 10 and
$10^4$. This indeed overlaps the range of Rm explored by our
simulations. 

Assuming a surface density of $0.02$ g/cm$^{-2}$  for $R\lesssim 10$
kpc, a typical scaleheight of 1000 parsecs,  and a rotation period of
200 Myrs, we find that the GI-dynamo produces $B\simeq 2\, \mu\text{G} -
5 \, \mu\text{G}$ for $\text{Rm} \sim 10-100$ in a handful of orbits. 
On the other hand, radio-faint galaxies like M31 and M33  possess fields strength  of about
5 $\mu\text{G}$, while gas-rich galaxies with high star-formation
rates, like M51, M83 and NGC 6946 possess $15\mu\text{G}$ on
average. We stress, however, that direct comparison is difficult and
relies on various, somewhat dubious, assumptions.  The value of the anomalous resistivity
found in the literature is  ad-hoc and estimated from simulations or
mean field models.   Moreover, our simulations do not describe the rich
assortment of physical processes occurring in galactic disc gas, such
as supernova driven-turbulence or other stellar feedback. Simulations taking
into account these effects are necessary to assess their relative
importance in comparison to GI. Other than differences in thermodynamics, one of the more
glaring discrepancies is our use of a Keplerian rotation law which is, of course, 
inappropriate for galaxies, though easily rectified in future simulations. Finally,
our local model cannot describe the development of the bar instability, which itself may 
impact significantly (positively or negatively) on a galaxy’s magnetic field generation.   

\section{Conclusions}
\label{sec_conclusions}

In summary, we performed 3D MHD stratified shearing box simulations,
with zero-net magnetic flux, in order to assess the ability of gravito-turbulence
to sustain a dynamo in the presence of finite and uniform
resistivity. In Section \ref{sec_threshold} and
\ref{sec_dynamo_scale}, we explored the fundamental properties of
the dynamo by probing a wide range of magnetic Reynolds numbers,
cooling times, and boundary conditions. We obtained the following
results: 
\begin{enumerate}[(a)]
\item The dynamo works for cooling times
  between $\tau_c=5$ (which corresponds roughly to the fragmentation
  threshold) and $100\, \Omega^{-1}$.  Growth rates and saturated
  magnetic energy decrease with  $\tau_c$. \\
\item The dynamo is robust to boundary conditions, 
   but its properties depends on which boundary condition is applied vertically (periodic or open).\\
\item In the case of open boundaries, the critical magnetic Reynolds
  number is of order unity ($\text{Rm}_c\simeq 4$ for
  $\tau_c=\,20\Omega^{-1}$). It is around 200 when periodic boundary
  conditions are used. \\
\item Magnetic energy reaches quasi-equipartition with kinetic
  pressure ($\beta_t=1$) for $\text{Rm}\lesssim 100$. For larger Rm
  however, the dynamo growth rate and saturated magnetic energy drop
  with Rm, which suggests that the dynamo is ``slow". {However this last statement has to be taken with caution since it is not possible to predict the dynamo behavior in the large magnetic Reynolds number limit.}  \\
\item The  dynamo is essentially large scale, kinematic and of mean
  field type, although smaller magnetic scales emerge at large Rm.
  It is powered by both differential rotation and large-scale
  spiral waves motion. 
\end{enumerate}

In Section \ref{sec_spirals},  we investigated the role of spiral
waves in the magnetic field amplification and proposed a model inspired by the solar dynamo cycle and based on the classical
$\alpha$-$\Omega$ theory of \citet{parker55}. This model assumes that:
\begin{enumerate}[(a)]
\item  Horizontal motions associated with spiral waves compress an
  initial toroidal field along the density waves.
\item Pairs of vertical counter-rotating
  rolls (associated with the density wave) pinch the field, lift, and
  fold it, generating new radial field.
\item The disk's differential rotation shears out the radial field to
  regenerate toroidal field, and the loop is closed.
\end{enumerate}
We demonstrated that the vertical EMF profiles giving rise
to the large-scale field are generated by the rolls. 
At large Rm, we found however that the large-scale
magnetic ropes inside the spiral waves are braided and broken by
small-scale parasitic modes, probably associated with a kink
instability.  Small-scale fast reconnection processes, like the
tearing instability, might also be involved in the destruction of the
large scale field at large Rm.

As discussed in Section \ref{sec_discussion}, the GI-dynamo  could have
direct application to young and massive gravito-turbulent stellar
discs, since these objects are cold and exhibit
large resistivities. In the coming years, ALMA will have the potential
to directly infer magnetic field orientation by measuring  Zeeman
splitting of cyanide lines. This technique has been already utilised
to explore
large-scale massive objects such as molecular clouds, but is now
applicable to circumstellar discs  \citep{brauer17}.   This raises
the hope of detecting spiral magnetic patterns in PP discs and therefore
of characterizing their dynamo behaviour. The highly resistive 
dead zones of older PP disks undergoing
outbursts are another venue in which the GI dynamo phenomena might
appear, possibly seeding the field necessary to ``kickstart'' the MRI.
More speculatively, our work might be applied to galactic dynamos,
which so far have only been addressed within the framework of mean
field theory. Our main result is that spiral waves can directly
produce dynamo action in these objects as well as the observed
alignment between magnetic and density spiral patterns {This result is in particular reminiscent of galactic dynamo simulations by \citet{Khoperskov18}}. So too, might the dynamo work in the cooler outer
radii of AGN disks.

Finally, we emphasise that characterising  the
behaviour of the dynamo at Rm larger than a thousand is not yet
possible with our simulations, because of resolution issues. This
potentially limits application of the theory to more astrophysical settings.
In addition, future work will be necessary to
better understand  the effect not only of low resistivity but also other
non-ideal effects, ambipolar diffusion especially. 
Discs are also generally threaded
by a net vertical field; the impact of such large scale component
remains unclear and will be treated separately.

\bibliographystyle{mnras}
\bibliography{refs} 




%
%
%
%
%

\label{lastpage}
\end{document}